\documentclass[twocolumn,preprintnumbers,amsmath,amssymb]{revtex4}
\usepackage{graphicx,color,psfrag}
\usepackage{amsmath,amssymb,latexsym}
\usepackage{bbm}

\newcommand{\Prob}{\mathbb{P}}

\newcommand{\Real}{\mathbb{R}}
\newcommand{\Att}{\mathbb{E}}

\newcommand{\X}{\{X_t\}_{t=1}^{\infty}}
\newcommand{\Y}{\{Y_t\}_{t=1}^{\infty}}
\newcommand{\I}{\{I_t\}_{t=1}^{\infty}}

\newcommand{\PDFpaper}{23}
\newcommand{\rhopaper}{20}
\newcommand{\EQPDFYpaper}{14}
\newcommand{\PDFYpaper}{13}
\newcommand{\phipaper}{15}
\newcommand{\mpaper}{27}

\begin{document}

\title{Scaling symmetry, renormalization, and time series modeling}

\author{Marco Zamparo}
\email{marco.zamparo@hugef-torino.org}
\affiliation{
HuGeF, Via Nizza 52, 10126 Torino, Italy}

\author{Fulvio Baldovin}
\email{baldovin@pd.infn.it}
\affiliation{
Dipartimento di Fisica e Astronomia, Sezione INFN, CNISM, 
Universit\`a di Padova,
 Via Marzolo 8, I-35131 Padova, Italy
}

\author{Michele Caraglio}
\email{caraglio@pd.infn.it}
\affiliation{
Dipartimento di Fisica e Astronomia, Sezione INFN, CNISM, 
Universit\`a di Padova,
 Via Marzolo 8, I-35131 Padova, Italy
}

\author{Attilio L. Stella}
\email{stella@pd.infn.it}
\affiliation{
Dipartimento di Fisica e Astronomia, Sezione INFN, CNISM, 
Universit\`a di Padova,
 Via Marzolo 8, I-35131 Padova, Italy
}

\date{\today}

\begin{abstract}
We present and discuss a stochastic model of financial assets dynamics
based on the idea of an inverse renormalization group strategy. With
this strategy we construct the multivariate distributions of
elementary returns based on the scaling with time of the probability
density of their aggregates. In its simplest version the model is the
product of an endogenous auto-regressive component and a random
rescaling factor designed to embody also exogenous
influences. Mathematical properties like increments' stationarity and
ergodicity can be proven.  Thanks to the relatively low number of
parameters, model calibration can be conveniently based on a method of
moments, as exemplified in the case of historical data of the S\&P500
index. The calibrated model accounts very well for many stylized
facts, like volatility clustering, power law decay of the volatility
autocorrelation function, and multiscaling with time of the aggregated
return distribution.  In agreement with empirical evidence in finance,
the dynamics is not invariant under time reversal and, with suitable
generalizations, skewness of the return distribution and leverage
effects can be included.  The analytical tractability of the model
opens interesting perspectives for applications, for instance in terms
of obtaining closed formulas for derivative pricing.  Further
important features are: The possibility of making contact, in certain
limits, with auto-regressive models widely used in finance; The
possibility of partially resolving the long-memory and short-memory
components of the volatility, with consistent results when applied to
historical series.
\end{abstract}

\maketitle

\section{Introduction}

Time series analysis plays a central role in many disciplines, like
physics \cite{kantz}, seismology \cite{shcherbakov}, biology
\cite{wilkinson}, physiology \cite{ivanov}, linguistics
\cite{petersen}, or economy \cite{preis}, whenever datasets amount to
sequences of measurements or observations.  
A main goal of such analysis is that of capturing
essential regularities of apparently unpredictable signals within a
synthetic model, which can be used to get forecasts and a deeper
understanding of the mechanisms governing the processes under study.
A satisfactory time series modeling for complex systems may become
a challenging task, due to the need to account for statistical
features of the data connected with the presence of strong
correlations.  In the last decades, features of this kind have been
extensively studied in the context of financial time series, where
they strongly stimulated the search for adequate stochastic modeling
\cite{bouchaud_1,tsay_1,musiela_1}.  The non-Gaussianity of the
probability density function (PDF) of aggregated returns of an asset
over time intervals in substantial ranges of scales, its anomalous
scaling and multiscaling with the interval duration, the long-range
dependence of absolute return fluctuations (volatility), the violation
of time-reversal symmetry, among other robust statistical features
called {\it stylized facts} in finance \cite{cont_1,cont_3}, still
remain elusive of synthetic and analytically tractable modeling.
Besides the standard model of finance
based on geometric Brownian motion \cite{bouchaud_1,voigt}, 
proposed descriptions include
stochastic volatility models (See, e.g., \cite{musiela_1} and
references therein), multifractal models inspired by turbulence
\cite{ghashghaie_1,mandelbrot_1,vassilicos_1,bacry_1,eisler_1,borland_1},
multi-timescale models \cite{zumbach_1,borland_2}, various types of
self-similar processes
\cite{mantegna_1,mantegna_2,baldovin_2,stella_1,peirano_1,andreoli_1},
multi-agent models \cite{lux_1,lebaron_1,alfi_1}, and those in the
Auto-Regressive Conditional Heteroskedastic (ARCH) and
Generalized-ARCH (GARCH) family
\cite{engle_1,bollerslev_2,bollerslev_1,tsay_1}.

To be effective, a stochastic model should not only correctly
reproduce the statistical features observed in the empirical analysis,
but also be easy to calibrate and analytically tractable in order to
be useful in applications like derivative pricing and risk evaluation
\cite{bouchaud_1,hull_1}.  In this respect, research in stochastic
modeling of financial assets is still a challenging topic
\cite{borland_1}.  Recently, some of us proposed an approach to market
dynamics modeling \cite{baldovin_2,baldovin_1} inspired by the
renormalization group theory of critical phenomena
\cite{kadanoff_1,lasinio_1,goldenfeld_1}.  The background ideas
exposed in Refs. \cite{baldovin_2,baldovin_1} already stimulated some
contributions along various lines
\cite{baldovin_2,stella_1,baldovin_3,baldovin_4,baldovin_5,peirano_1,andreoli_1}.
In particular, in \cite{peirano_1} a model with nonstationary
increments and lacking a volatility feedback mechanism has been
discussed in detail, pointing out its potential interest and missing
features.  In the present Paper, we make a step forward
along the lines proposed in \cite{baldovin_2}, by introducing a novel
discrete-time stochastic process characterized by both an
auto-regressive endogenous component and a short-memory one.
The firt provides a volatility feedback
thanks to its long dynamical memory; 
the latter represents, besides 
immediate endogenous mechanisms, also the
impact of external influences.  Many features of the model are under
analytical control and a number of basic properties, like increments'
stationarity, time-reversal asymmetry, strong mixing and ergodicity as
a consequence, can be proved.  In addition, an explicit procedure for
calibrating its few parameters makes the model a
candidate for applications, e.g., to derivative pricing
\cite{bouchaud_1,hull_1}, for which useful closed expressions can be
derived \cite{baldovin_6}.  An interesting feature of our approach is
the possibility of resolving the long-memory and short-memory
components of the volatility. This could be exploited
in order to partially separate exogenous and endogenous mechanisms
within the market dynamics.  
The version of the model we discuss within the
present Paper does not include skewness in the return PDF and the
leverage effect \cite{bouchaud_1,bouchaud_lev}. However, here we
outline simple ways of improving it in order to consistently recover
also these effects.

While some analytical derivations and detailed proofs are reported in
the Supplementary Material \cite{supplementary}, in the main text we
illustrate the general ideas inspiring the model, we discuss its
properties, and show that they allow to implement a successful
calibration protocol.  Specifically, we use the model to reproduce the
daily historical time series of the S\&P500 index in the years
1950-2010.  The Paper is organized as follows. The next Section
contains a description of the background ideas inspiring the model
construction, whose precise definition is reported in Section
\ref{sec_construction}.  This Section also describes a simple
parametrization in which contact with a Markov-switching ARCH process
(SWARCH) is realized.  Section \ref{sec_modelproperties} is then
devoted to a brief review of the properties of the model.  Section
\ref{sec_calibration} proposes a simple calibration scheme, whereas a
comparison with historical series is discussed in Section
\ref{sec_main}.  Section \ref{sec_endex} deals with the interesting
question about identifying the long-memory and short-memory components
in empirical time series.  In Section \ref{sec_improvements} we
discuss the possibility of describing skewness and the leverage effect
and mention other perspective developments.  Finally, in Section
\ref{sec_conclusions} we draw our conclusions.

\section{Scaling as a guiding symmetry}
Since the pioneering work of Mandelbrot and van Ness on fractional
Brownian motion \cite{mandelbrot_4}, interest in scaling features has
characterized many models of financial and other time series,
especially in contributions by members of the physics community.
Proposals include the representation of financial processes as
truncated Levy flights \cite{mantegna_1,mantegna_2}, or the more
sophisticated descriptions through multifractal cascades inspired by
turbulence
\cite{ghashghaie_1,mandelbrot_1,vassilicos_1,bacry_1,eisler_1,borland_1}.
In the financial literature, a similar focus on scaling properties is
harder to find.  Indeed, although leptokurtic distributions of
aggregated returns are typically obtained in ARCH and similar models
by making the conditioned variance of successive elementary increments
dependent on the past history
\cite{engle_1,bollerslev_2,bollerslev_1,tsay_1}, even for more
specialized versions of this type of approach, like FIGARCH
\cite{baillie_1}, a proper description of the correct scaling and
multiscaling properties of aggregated increments is still an open
issue.
  
A cornerstone achievement in statistical physics has been the
formulation of the renormalization group approach to scaling
\cite{kadanoff_1,lasinio_1,goldenfeld_1}.  In this approach one tries
to deduce the scaling properties of a system at criticality by
analyzing how coarse-graining operations, which rescale the minimal
{\it length} at which the system is resolved, change its statistical
description in terms of effective interactions or similar
parameters. The scaling invariance at criticality then emerges when
such changes do not occur (fixed point).  In a recent publication
\cite{baldovin_2}, some of the present authors made the proposal that
the problem of modeling the stochastic financial process on {\it time
  scales} for which a well defined scaling symmetry holds at least
approximately, may be faced by inverting the logic of the standard
renormalization group procedure.  Given as an input the scaling
properties of the aggregated increment PDF over a certain time scale,
the idea is to find by fine-graining basic probabilistic rules that
apply to elementary increments in order for them to be consistent with
the input scaling properties \cite{stella_1,baldovin_1}. These rules
should describe the system on time scales shorter than that of the
aggregation interval, and their knowledge is regarded to be equivalent
to that of the effective fixed point interactions in the standard
renormalization group approach. Of course, even if properties like the
martingale character of the financial process pose strong constraints,
there is a degree of arbitrariness in the fine graining operation, and
an important task is to show that the proposed fine-graining is
plausible at the light of the relevant stylized facts.

This fine-graining, reverse renormalization group strategy for the
description of market dynamics has been already exemplified in
previous contributions \cite{baldovin_2,peirano_1,andreoli_1},
especially dealing with high frequency processes
\cite{baldovin_3,baldovin_4,baldovin_5}.  Unlike in cases for which a
single time series is available, in
Refs. \cite{baldovin_3,baldovin_4,baldovin_5} we focused on a
particular, fixed window of the daily evolution of an asset, and
extracted from the available records an ensemble of histories which
have been assumed to be independent realizations of the same
stochastic process. The manifest time inhomogeneous character of this
process and its limited duration in time significantly simplify a
modeling approach based on the above fine-graining strategy.  Things
become more difficult when only one realization of the process one
wishes to model is available in the form of a single, long time
series.  This is the situation we discuss in the present work.

While a precise mathematical definition of our model is postponed to
the next Section, in the present one we summarize the basic ideas
behind its construction. In particular, we emphasize the inspiration
by the renormalization group approach and the basic complementary role
played by both endogenous and exogenous mechanisms.  Another key
aspect concerns the introduction of an auto-regressive dynamical
scheme. In our context this endows the endogenous mechanism with
sufficiently long memory, guaranteeing at the same time strong mixing,
and hence also the ergodicity of the process \cite{mathpaper}.

Let $\{X_t\}_{t=1}^\infty$ be a sequence of random variables
representing the increments (logarithmic returns in finance) of a
discrete-time stochastic process. This process possesses a
simple-scaling symmetry if $X_1+\cdots+X_t$ has the same probability
law as $t^H X_1$ for any $t$, $H>0$ being the scaling (Hurst)
exponent. If this is the case, the property
\begin{equation}
t^H\;g_t(t^H\;x)=g(x)
\label{coll}
\end{equation}
holds for the PDF $g_t$ of the aggregated increments $X_1+\cdots+X_t$, 
where $g$ is the scaling function (which also coincides with the PDF of
$X_1$).  
One immediate consequence of Eq.\ (\ref{coll}) is a
scaling property for the existing moments of the process:
\begin{equation}
\mathbb E[|X_1+\cdots+X_t|^q]=t^{q\,H}\;\mathbb E[|X_1|^q].
\label{mscaling}
\end{equation}
A normal scaling symmetry is obtained with $g$ Gaussian and $H=1/2$.
Anomalous scaling refers to the fact of $g$ not being Gaussian and/or
$H\neq1/2$. Another kind of anomalous behavior for which
Eq.\ (\ref{mscaling}) holds with an exponent $H_q$ explicitly
depending on the moment order $q$ is called {\it multiscaling} and in
this case $H_q$ is also named generalized Hurst exponent
\cite{frisch_1,di_matteo_1}.

The simple-scaling symmetry can also be expressed in terms of
characteristic functions (CF) as
\begin{equation}
\mathbb{E}\bigl[e^{i k(X_1+\cdots +X_t)}\bigr]=\mathbb{E}\bigl[e^{i k (t^H X_1)}\bigr],
\end{equation}
or, equivalently, as
\begin{equation}
\widehat f_t^{\,X}\left(k,\ldots,k\right)=\widehat f_1^{\,X}\left(t^H k\right),
\label{eq_scaling_cf_joint}
\end{equation}
where $\widehat f_t^{\,X}\left(k_1,\ldots,k_t\right)\equiv
\mathbb{E}\bigl[e^{i(k_1X_1+\cdots +k_t X_t)}\bigr]$ is the joint CF
of $X_1,\ldots,X_t$, i.e.\ the Fourier transform of the joint PDF
$f_t^{X}(x_1,\ldots,x_t)$.

Aiming at constructing a model for the increments consistent with
Eq.\ (\ref{coll}) for a given scaling exponent $H>0$ and general
scaling function $g$, we notice that the knowledge of $f_1^{X}=g$
combined with Eq.\ (\ref{eq_scaling_cf_joint}) allows us to only fix
the CF $\widehat f_t^{\,X}$ along the diagonal:
\begin{equation}
\widehat f_t^{\,X}\left(k,\ldots,k\right)=\widehat g\left(t^H k\right)\equiv \int_{\mathbb{R}} d x\;e^{i (t^H k) x}\;g(x).
\end{equation}
The basic inspiration of our approach is thus a quest for the
existence of conditions implied by the presence of anomalous scaling
which allow us to determine this joint CF also off-diagonal. Namely:
``{\it Are there ways of fixing $\widehat
  f_t^{\,X}\left(k_1,\ldots,k_t\right)$ such that $\widehat
  f_t^{\,X}\left(k,\ldots,k\right)=\widehat g(t^H k)$ with $g$
  non-Gaussian and/or $H\neq 1/2$ assumed to be given?}''  As a rule,
when applying the renormalization group approach, one would be faced
with the inverse problem: given a parametric form for $\widehat
f_t^{\,X}$, or for $f_t^X$, one tries to fix its parameters in such a
way that Eq.\ (\ref{eq_scaling_cf_joint}), and thus Eq.\ (\ref{coll})
with $H$ and $g=f_1^X$ to be determined, is satisfied. This amounts to
the identification of the fixed point and is generally accomplished by
operating a suitable coarse-graining operation on the description of
the process. The fixed point is just an instance of the process which
is left invariant under such operation.  To satisfy our quest, we need
to implement a plausible inversion of the coarse-graining operation in
which the fixed point scaling is assumed to be known and $\widehat
f_t^{\,X}$ needs to be constructed. This inverse procedure is not
unique in general and its plausibility needs to be tested {\it a
  posteriori}. We are thus somehow ``reverting'' the ordinary flux in
a renormalization-group approach, as we are trying to realize a {\it
  fine-graining} procedure compatible with the existence of an
anomalous fixed point scaling.

Given $H$ and $g$ as an input, our proposal is to set
\begin{equation}
\widehat f_t^{\,X}\left(k_1,\ldots,k_t\right)
=\widehat g\left(\sqrt{a_1^2\;k_1^2+\cdots+a_t^2\;k_t^2}\right)
\label{eq_ftil}
\end{equation}
where
\begin{equation}
a_i=
\sqrt{i^{2H}-(i-1)^{2H}}
\label{eq_ai}
\end{equation}
for any $i\in\mathbb N^+$, and to find conditions on $g$ which
guarantee that such a $\widehat f_t^{\,X}$ is a proper CF. If this is
the case, $\widehat f_t^{\,X}$ is the Fourier transform of a PDF and
manifestly solves Eq.\ (\ref{eq_scaling_cf_joint}).  We thus meet with
the problem of characterizing the class of scaling functions $g$ which
make the inverse Fourier transform of our trial CF a non-negative
joint PDF. Fortunately, this problem is addressed by Schoenberg's
theorem \cite{schoenberg_1,aldous_1}, which guarantees that
Eq.\ (\ref{eq_ftil}) provides a proper CF for all $t$ if and only if
$\widehat g$ is of the form
\begin{equation}
\widehat g(k)
=\int_{0}^\infty d\sigma
\;\rho(\sigma)
\;e^{-\sigma^2\,k^2/2},
\end{equation}
$\rho$ being a PDF on the positive real axis.  The class of scaling
functions suitable for our fine-graining procedure is thus constituted
by the Gaussian mixtures
\begin{equation}
g(x)
=\int_{0}^\infty d\sigma
\;\rho(\sigma)
\;\mathcal N_\sigma(x),
\label{eq_scaling_function_g}
\end{equation}
where, here and in the following, $\mathcal N_\sigma$ denotes a
Gaussian PDF with mean zero and variance $\sigma^2$. Such a class,
whose elements are identified by $\rho$, is rich enough to allow us to
account for very general anomalous scaling symmetries.  The joint PDF
of the variables $X_t$'s provided by our inverse strategy and
corresponding to $g$, i.e.\ to $\rho$, is then obtained by applying an
inverse Fourier transform to Eq.\ (\ref{eq_ftil}) and reads
\begin{equation}
f_t^X(x_1,\ldots,x_t)=
\int_{0}^\infty d\sigma
\;\rho(\sigma)
\;\prod_{i=1}^t
\;\mathcal N_{a_i\;\sigma}(x_i),
\label{eq_mixture}
\end{equation}
with the $a_i$'s as in Eq.\ (\ref{eq_ai}).  

There are various ways in which Eq.\ (\ref{eq_mixture}) can inspire
the construction of a stochastic process suitable for finance.  Some
possibilities have been tested in
Refs.\ \cite{stella_1,peirano_1,andreoli_1,baldovin_3,baldovin_4}.  In
general, the joint PDF of Eq.\ (\ref{eq_mixture}) itself cannot
describe a stationary ergodic sequence $\{X_t\}_{t=1}^\infty$, but for
the problem we address here, i.e. to describe long historical time
series, such features are relevant. To recover 
stationarity and ergodicity keeping contact with Eq.\ (\ref{eq_mixture}), we here
conceive the process of the returns as separated into two
components. As shown below, the manifest scaling property of the
Gaussian
\begin{equation}
\mathcal N_{a\,\sigma}(x)=\frac{1}{a}\;\mathcal N_{\sigma}\left(\frac{x}{a}\right),
\end{equation} 
which holds for any $a>0$, prompts such a separation.  In the
financial time series context, one is then naturally led to interpret
these components as accounting for long-memory endogenous dynamical
mechanisms and for the occurrence of short-memory endogenous and
exogenous events, respectively.

As far as the former component is concerned, a correlated process
$\{Y_t\}_{t=1}^\infty$ with memory order $M$ is considered.  Up to $t$
equal to $M$, this process is characterized by the joint PDF of
Eq.\ (\ref{eq_mixture}) with $a_i=1$ for all $i$'s.  This is a
sequence of non-Gaussian, dependent random variables and at times up
to $M$ their sum satisfies a form of anomalous scaling with $H=1/2$
and $g$ given by Eq.\ (\ref{eq_scaling_function_g}).  The introduction
of a finite $M$ of course limits the range of time for which this form
of scaling is valid.  This is not a problem, because empirically we
know that anomalous scaling approximately holds within a finite time
window.  The entire process $\{Y_t\}_{t=1}^\infty$ is then obtained
through an auto-regressive scheme of order $M$.  This auto-regressive
scheme is such to prevent the dynamics from stabilizing the
conditional variance of $Y_t$'s, given the past history, to a constant
value after an initial transient, thus restoring full ergodicity
\cite{stella_1,mathpaper}.  At the same time, the conditioning effect
of the previous values of the process on the future dynamical
evolution is of primary importance in applications like, e.g., those
related to derivative pricing \cite{baldovin_6} or volatility
forecasts.

The latter component introduces a multiscaling behavior by multiplying
each element of the above 
sequence by the corresponding factor $a_t$ given in
Eq.\ (\ref{eq_ai}): $X_t=a_t Y_t$.  In principle, these rescaling
factors convey a time inhomogeneity to the increments $X_t$'s, a
property which has been exploited in the modeling of ensembles of
histories \cite{baldovin_3,baldovin_4,baldovin_5}.  However, a proper
randomization of the time argument of the $a_t$'s 
restores the stationarity of the $X_t$'s, making them
suitable for describing single time series whose statistical
properties are thought to be independent of time \cite{cont_1}.  This
randomization, which is obtained by the introduction of a
  short-memory process, is regarded as mimicking the effects on
market evolution of both short-memory endogenous random factors
and external inputs of information or changing conditions, thus
conferring also an exogenous character to this second component. The
first component, which is responsible for the volatility clustering
phenomenon thanks to its possible long dynamical memory $M$, is
then interpreted as the long-memory endogenous part.  In order to
  have a simple intuition of the returns' compound process, we may
  sketch a comparison with electronics and telecommunications
  regarding the long-memory component as a carrier signal, which is
  modulated by the short-memory one, playing thus the role of a
  modulating signal.

As we shall review in the Paper, and show in the Supplementary
Material \cite{supplementary}, relevant properties of the model, like
its multiscaling and the power-law decay of non-linear
autocorrelations over finite time horizons, are determined by
the short-memory component.  We stress that when combining the
long-memory and short-memory processes, together with simple scaling
features also the direct link between the Hurst exponent and the
exponent $H$ entering in Eq.\ (\ref{eq_ai}) is lost. For this reason,
in the following we will denote by $D$, instead of $H$, the parameter
involved in the definition of the model
[See Eq. (\ref{eq_a_i}) below].

\section{Model definition}
\label{sec_construction}
On the basis of the background material elaborated in the previous
Section, here we precisely define our stochastic process of the
increments.  Such a process $\{X_t\}_{t=1}^\infty$ is obtained as the
product of an endogenous auto-regressive component
$\{Y_t\}_{t=1}^\infty$ and a rescaling, or modulating, factor
$\{a_{I_t}\}_{t=1}^\infty$, where $\{I_t\}_{t=1}^\infty$ is a discrete
Markovian random time independent of $\{Y_t\}_{t=1}^\infty$, and
$\{a_i\}_{i=1}^\infty$ is a positive sequence:
\begin{equation}\label{eq_processX}
X_t\equiv  a_{I_t}Y_t.
\end{equation}

The stochastic process $\{Y_t\}_{t=1}^\infty$ is a Markov process
taking real values with memory $M>0$. It is defined, through its
PDF's, by the following scheme:
\begin{equation}
f_t^Y(y_1,\ldots,y_t)
\equiv \varphi_t(y_1,\ldots,y_t)
\label{gp}
\end{equation} 
if $t=1,2,\ldots,M$, and
\begin{eqnarray}
f_t^Y(y_1,\ldots,y_t)
&\equiv &\frac{\varphi_{M+1}(y_{t-M},\ldots,y_t)}{\varphi_M(y_{t-M},\ldots,y_{t-1})}\;\cdot
\nonumber\\
&\cdot& f_{t-1}^Y(y_1,\ldots,y_{t-1})
\label{gs}
\end{eqnarray} 
if $t>M$. Here, the PDF's $\varphi_t$ are given by
\begin{equation}
\varphi_t(y_1,\ldots,y_t)
\equiv 
\int_{0}^\infty d\sigma
\;\rho(\sigma)
\;\prod_{n=1}^t
\;\mathcal N_\sigma(y_n).
\label{gp_1}
\end{equation}

The process $\{I_t\}_{t=1}^\infty$ is a Markov chain of order 1 valued
in $\mathbb N^+$.  The memory order 1 of this sequence, to be compared
with the memory order $M$ of the above one, justifies our convention of
referring the two components as ``short-memory'' and ``long-memory'', 
respectively.  The chain $\{I_t\}_{t=1}^\infty$ is
defined by the initial condition
\begin{equation}
  \mathbb P[I_1=i]\equiv \nu(1-\nu)^{i-1}
  \label{eqi}
\end{equation}
and by the transition probabilities 
\begin{equation}
  \mathbb P[I_{t+1}=i|I_t=j]\equiv 
  \begin{cases}
    \nu & \mbox{if }i=1;\\
    1-\nu & \mbox{if }i=j+1;\\
    0 & \mbox{otherwise}.
  \end{cases}
  \label{Wi}
\end{equation}
In words, we are stating that at time $t+1$ there is a ``time-reset''
or ``restart'' ($I_{t+1}=1$) with probability $\nu>0$, whereas with
probability $1-\nu$ time flows normally ($I_{t+1}=I_t+1$). For
notational simplicity we set \mbox{$\pi(i)\equiv \mathbb P[I_1=i]$} and we
collect the transition probabilities into a stochastic matrix with
entries \mbox{$W(i,j)\equiv \mathbb P[I_{t+1}=i|I_t=j]$}. We point out that
our choice of $\pi$ corresponds to the invariant distribution of $W$,
with the consequence that $\{I_t\}_{t=1}^\infty$ turns out to be a
stationary process:
\begin{equation}
\sum_{j=1}^{\infty}W(i,j)\pi(j)=\pi(i).
\end{equation}
It should be stressed that here we assume that $\{Y_t\}_{t=1}^\infty$
and $\{I_t\}_{t=1}^\infty$ are independent in favor of an initial
simplicity. As a consequence, the present model results in a
Markov-switching model where, by definition, the switching mechanism
between different regimes 
is controlled by an unobservable state variable that
follows a first-order Markov chain.  In Section \ref{sec_improvements}
we shall then hint at the possibility of making the random time
$\{I_t\}_{t=1}^\infty$ dependent on $\{Y_t\}_{t=1}^\infty$.

Finally, $\{a_i\}_{i=1}^\infty$ is a positive sequence where,
without loss of generality,  
we can set $a_1=1$.   
In analogy with the previous Section, we assume a factor $a_i$ of the
form
\begin{equation}
\label{eq_a_i}
a_i=\sqrt{i^{2D}-(i-1)^{2D}}
\end{equation}
with $D>0$. The relation between the Hurst exponent and the model
parameter $D$ will be addressed in what follows.  For the moment, let
us point out that the sequence $a_i$ is identically equal to 1 if
$D=1/2$ while monotonically decays to zero or diverges if $D<1/2$ or
$D>1/2$, respectively. For financial applications, the instance
$D<1/2$ appears to be the interesting one and, since
$\lim_{i\to\infty}i^{1/2-D}a_i=\sqrt{2D}$, the decay of the rescaling
factor is of power-law type.  However, in principle other choices for
the functional form of $\{a_i\}_{i=1}^\infty$ are possible and could
be introduced for further extensions and applications of the model.

The endogenous process $\{Y_t\}_{t=1}^\infty$ recalls the ARCH
construction of order $M$ \cite{engle_1} because the conditional PDF
of the current $Y_t$, given the past history, depends on the previous
outcomes only through the sum of the squares of the latest $M$ ones,
as one can easily recognize.  As a matter of fact,
$\{Y_t\}_{t=1}^\infty$ becomes a genuine ARCH process if the function
$\rho$ is properly chosen, as we shall show in a moment. In general, the
basic difference with respect to an ARCH process is that
here the whole conditional PDF of $Y_t$, and not only its variance,
changes with time.  In spite of this, the process
$\{Y_t\}_{t=1}^\infty$ is identified by a small number of parameters
independently of the order $M$. Indeed, 
besides $M$, the parameters associated to
$\{Y_t\}_{t=1}^\infty$ are only those related to $\rho$.  As we
discuss below, satisfactory parametrizations of $\rho$ for financial
time series require just two parameters.  This must be contrasted with
the fact that in realistic ARCH models the number of
parameters can proliferate with the memory, easily becoming of the
order of several tens \cite{tsay_1}. Such a synthetic result, which we believe to be
a most interesting innovative feature of $\{Y_t\}_{t=1}^\infty$, is
made possible by the exploitation of the scaling symmetry embodied in
Eqs.\ (\ref{gp}--\ref{gp_1}).

A most practical choice for $\rho$ is one which allows us to
explicitly perform the integration over $\sigma$ in
Eq.\ (\ref{gp_1}). Indeed, we notice that weighing $\sigma^2$
according to an inverse-gamma distribution is the way to reach this
goal. Furthermore, in the context of financial modeling, this
prescription is in line with the rather common belief that the
distribution of the square of the empirical returns can be modeled as
an inverse-gamma distribution
\cite{gerig_1,micciche_1,peirano_1}. This $\rho$ is identified by two
parameters, $\alpha$ and $\beta$ governing its form and the scale of
fluctuations, respectively, and reads
\begin{equation}
\rho(\sigma)=\frac{2^{1-\frac{\alpha}{2}}}{\Gamma(\frac{\alpha}{2})}
\;\frac{\beta^\alpha}{\sigma^{\alpha+1}}
\text{e}^{-\frac{\beta^2}{2\sigma^2}},
\label{rhoinv}
\end{equation}
where $\Gamma$ denotes the Euler's gamma function.  Interestingly,
making this choice within the model, the endogenous component
$\{Y_t\}_{t=1}^{\infty}$ becomes a true ARCH process of order $M$ with
Student's t-distributed return residuals, as anticipated
above. Indeed, in the Supplementary Material \cite{supplementary} we
prove that if $\rho$ is given by Eq.\ (\ref{rhoinv}), then we can
reformulate our model as $X_t=a_{I_t}Y_t$ with
\begin{equation}
Y_t=\begin{cases}
\beta\cdot Z_1 & \mbox{if } t=1;\\
\sqrt{\beta^2+\sum_{n=1}^{\min\{t-1,M\}}Y_{t-n}^2\;}\cdot Z_t & \mbox{if } t>1,
\end{cases}
\end{equation}
and the return residual process $\{Z_t\}_{t=1}^{\infty}$ amounting to
a sequence of independent Student's t-distributed variables:
\begin{equation}
f_t^Z(z_1,\ldots,z_t)=\prod_{n=1}^t\frac{\Gamma(\frac{\alpha_n+1}{2})}{\sqrt{\pi}\;\Gamma(\frac{\alpha_n}{2})}
\;(1+z_n^2)^{-\frac{\alpha_n+1}{2}}
\label{Zdist}
\end{equation}
with $\alpha_n\equiv \alpha+\min\{n-1,M\}$. It is also worth noticing
that the Markov-switching character of the volatility, introduced by
the process $\{a_{I_t}\}_{t=1}^{\infty}$, reconciles this particular
instance of our model with the SWARCH category proposed by Hamilton
and Susmel \cite{hamilton_1}. The only difference, apart from dealing
with an infinite number of regimes corresponding to the infinite
possible values taken by $I_t$, is that these regimes never persist
for more than one time step.
We stress however that besides $M$ only two parameters, $\alpha$ and
$\beta$, are here needed to completely specify
$\{Y_t\}_{t=1}^{\infty}$.  This typically applies also to other
possible parametrizations of $\rho$, not related to the inverse-gamma
distribution.

The $\{a_{I_t}\}_{t=1}^{\infty}$ component entails our model with two
further parameters, i.e.\ $\nu$ establishing the frequency of
occurrence of the ``time restarts'', and the exponent $D$ defining the
modulating factor $\{a_i\}_{i=1}^{\infty}$.  In summary, the model is
thus typically identified by 5 parameters, three related to the
long-memory and two to the short-memory components.  The general fact
that both $\{Y_t\}_{t=1}^{\infty}$ and $\{a_{I_t}\}_{t=1}^{\infty}$
are {\it hidden processes}, not separately detectable, complicates the
effectiveness of a parameter calibration protocol.  However, as
discussed below, analytical features of the model allow us to identify
moment optimization procedures that guarantee, for sufficiently long
time-series, proper determination of the input parameters.

In the next Section we clarify in details up to what extent the
scaling symmetry is preserved by the process
$\{X_t\}_{t=1}^\infty$. For the moment we point out that the contact
with the ARCH and Markov-switching models' literature is particularly
interesting.  Indeed, thanks to our general results below it sheds
some light on how to obtain anomalous scaling properties in
auto-regressive models on limited temporal horizons
\cite{mantegna_2}.

\section{Model properties}
\label{sec_modelproperties}
A number of properties of our model are independent of the choice of
the function $\rho$ and can be analytically investigated. Here we
briefly review these properties referring to the Supplementary
Material \cite{supplementary} for detailed derivations.

\subsection{Joint PDF and stationarity}
\label{sub_PDF}

For any $t\ge 1$ the joint PDF of $X_1,\ldots,X_t$ is given by the
formula
\begin{eqnarray}
f_t^X(x_1,\ldots,x_t)
&=&\sum_{i_{1}=1}^\infty\cdots\sum_{i_{t}=1}^\infty \; \prod_{n=1}^{t-1}W(i_{n+1},i_{n})\;\pi(i_1)\;\cdot
\nonumber\\
&\cdot&\frac{f_t^Y(x_1/a_{i_1},\ldots,x_t/a_{i_t})}{a_{i_1}\cdots a_{i_t}}.
\label{ft}
\end{eqnarray}
Since $f_t^Y$ is defined via mixtures of centered Gaussian variables,
Eq.\ (\ref{gp_1}), we realize immediately that the conditional
expectation of $X_t$, given the past history, vanishes. The process
$\{X_t\}_{t=1}^{\infty}$ is thus a martingale difference sequence,
reflecting the efficient market hypothesis \cite{fama_1,fama_2}.
Moreover, the structure of $f_t^X$ shows that the observed process
cannot retain the Markov property characterizing both 
$\{Y_t\}_{t=1}^{\infty}$ and  $\{a_{I_t}\}_{t=1}^{\infty}$, 
with the consequence that its
future evolution always depends on all past events. This feature
reflects the impossibility of directly detecting from the examination
of $\{X_t\}_{t=1}^{\infty}$ the random time $\{I_t\}_{t=1}^{\infty}$.
More importantly, the latter fact makes a maximum-likelihood
estimation of the model parameters very difficult because of the too
onerous computational work needed.  Thus, one is forced to refer to
some moment optimization procedure for settling this issue. For this
reason, in the next Section we shall propose a simple implementation
of a generalized method of moments. A procedure to identify the most
probable time restarts by means of the calibrated model, valuable for
some applications like, e.g., in option pricing, will be also
discussed in Section \ref{sec_endex}.

A remarkable feature of the joint PDF $f_t^X$ is that it does not
explicitly depend on the memory range $M$ at short time scales.
Indeed, when $t\leq M+1$ from Eqs.\ (\ref{ft}) and (\ref{gp}) we have
[notice that Eq.\ (\ref{gs}) gives $f_{M+1}^Y=\varphi_{M+1}$]
\begin{eqnarray}
f_t^X(x_1,\ldots,x_t)
&=&\sum_{i_{1}=1}^\infty\cdots\sum_{i_{t}=1}^\infty \; \prod_{n=1}^{t-1}W(i_{n+1},i_{n})\;\pi(i_1)\;\cdot
\nonumber\\
&\cdot&\int_{0}^{\infty}d\sigma\;\rho(\sigma)\;\prod_{n=1}^t\mathcal N_{a_{i_n}\,\sigma\,}(x_n).
\label{ftsimple}
\end{eqnarray}
This fact implies that models with different memory orders $M$ and
$M'>M$, and the same other parameters, cannot be distinguished by
looking at their features at times shorter than or equal to $M+1$.
Observe also that the Gaussian mixture structure provided by our
fine-graining strategy and the random nature of the factor redefining
the typical magnitude of the fluctuations is particularly clear in
Eq.\ (\ref{ftsimple}).

Our process is strictly stationary, meaning that
$(X_n,\ldots,X_{n+t-1})$ is distributed as $(X_1,\ldots,X_t)$ for any
$n\ge 1$ and $t\ge 1$. This property directly follows from the fact
that $\{Y_t\}_{t=1}^{\infty}$ and $\{I_t\}_{t=1}^{\infty}$ are both
stationary processes and, in particular, tells us that
Eqs.\ (\ref{ft}) and (\ref{ftsimple}) give the PDF of any string of
$t$ consecutive variables extracted from $\{X_t\}_{t=1}^{\infty}$. We
stress that stationarity is a basic assumption in time series
analysis, when one is forced to reconstruct the underlying stochastic
process on the basis of a single, possibly long, time series.

We also point out that the long-memory endogenous component
$\{Y_t\}_{t=1}^{\infty}$ is not only a stationary sequence, but even a
reversible one: the law of $(Y_t,Y_{t-1}\ldots,Y_1)$ is the same as
the law of $(Y_1,\ldots,Y_{t-1},Y_t)$ for any $t\ge 1$. In contrast,
the observed process $\{X_t\}_{t=1}^{\infty}$ is not reversible, being
such time-reversal symmetry broken by the short-memory component.  In
Section \ref{sec_main} we shall better analyze this feature of the
model, attempting to quantify the time-reversal asymmetry of
$\{X_t\}_{t=1}^{\infty}$.

The single-variable PDF, which is the same for any $X_t$ thanks to
stationarity, is obtained by setting $t=1$ in Eq.\ (\ref{ftsimple})
and explicitly reads
\begin{equation}
f_1^X(x)
=\sum_{i=1}^\infty\;\nu(1-\nu)^{i-1}
\int_{0}^{\infty}d\sigma\;\rho(\sigma)\;\mathcal N_{a_i\,\sigma\,}(x).
\label{eq_marginal_1}
\end{equation}
The mixture of Gaussian densities with different width can endow this
PDF with power law tails, as observed for financial assets
\cite{cont_1}.  Specifically, when $\rho(\sigma)$ decays as the
power-law $\sigma^{-\alpha-1}$ for large $\sigma$, $f_1^X$ becomes a
fat-tailed distribution with the same tail index $\alpha$. Thus, for
example choosing $\rho$ as in Eq.\ (\ref{rhoinv}) we have
$\lim_{x\to\infty}|x|^{\alpha+1}f_1^X(x)=c$ with
\begin{equation}
c\equiv \frac{\beta^\alpha\Gamma(\frac{\alpha+1}{2})}{\sqrt{\pi}\Gamma(\frac{\alpha}{2})}\sum_{i=1}^{\infty}\;a_i^{\alpha}\;\nu(1-\nu)^{i-1}<\infty,
\end{equation}
and the above form parameter $\alpha$ controls the tails of the PDF of
the $X_t$'s as long as $\nu$ is finite.  It is worth noticing that,
even if the above condition on $\rho$ is necessary for having fat
tails in a strict asymptotic sense, there is the possibility of
approximately realizing such a feature for returns in empirically
accessible ranges by only considering rare enough time restarts. As
explained in the Supplementary Material \cite{supplementary}, indeed,
assuming $\{a_i\}_{i=1}^{\infty}$ given by Eq.\ (\ref{eq_a_i}) with
$D<1/2$, and properly rescaling $\rho$ in order to avoid $f_1^X$ to
concentrate around zero in the small-$\nu$ limit, in general the
single-variable PDF displays fat tails with index $2/(1-2D)$ when the
restart probability $\nu$ approaches zero. Of course, with
$\rho(\sigma)$ behaving as $\sigma^{-\alpha-1}$ for large $\sigma$ and
$\alpha<2/(1-2D)$, the tail index is determined by $\alpha$ even in
the rare-restart limit.  In practice, when dealing with small values
of $\nu$ the empirically-accessible power law exponent of $f_1^X$
depends on all $\alpha$, $\nu$, and $D$.  This fact, and the
uncertainty affecting the empirical estimate of such exponent
\cite{cont_1}, lead us to a calibration protocol (See below) which is
not based on matching the effective power law tails of $f_1^X$.

Since we have here stated the stationarity of our model, we also
mention that strong mixing properties can be proved under mild
assumptions on the function $\rho$ \cite{mathpaper}. These mixing
properties entail ergodicity, which justifies the comparison between
empirical long time averages and theoretical ensemble expectations.
They also imply the validity of the central limit theorem, to which we
appeal for discussing scaling features of aggregated returns on the
long time horizon under, basically, the only hypothesis that the
second order moment of the elementary increments is finite.  Stating
precisely these results and discussing their proof however requires a
more rigorous setting \cite{mathpaper} which is beyond the scope of
the present Paper.  The ergodicity has also been numerically verified
on the basis of model-based simulations.

\subsection{Scaling features}
\label{sub_scaling}

\begin{figure}
\includegraphics[width=0.99\columnwidth]{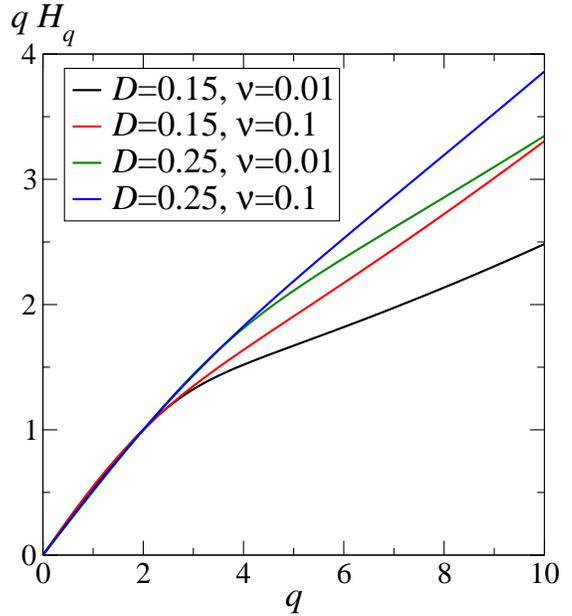}
\caption{Model multiscaling behavior for $1\leq t\leq31$ and $M\ge 30$.
  The couple $(D,\nu)$ determines the behavior.}
\label{fig_multiscaling_model}
\end{figure}

\begin{figure}
\includegraphics[width=0.99\columnwidth]{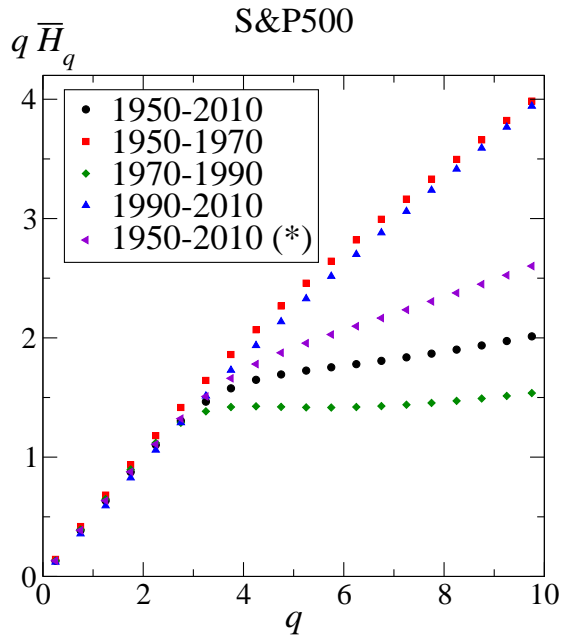}
\caption{ Simple- ($q\lesssim3$) and multi- ($q\gtrsim3$) scaling
  behavior of S\&P500 (log) returns PDF, analyzed for $t$ from 1 day to 2
  months in the years 1950-2010.  Notice the strong dependence of
  multiscaling features on the specific sample.  In particular, data
  marked with ($^*$) refer to the full interval 1950-2010 with an
  artificious alteration of a single data: a reduction of $20\%$ of
  the strongest fluctuation.  }
\label{fig_multiscaling_sep}
\end{figure}

Scaling features of $\{X_t\}_{t=1}^\infty$ are at the heart of our
approach and two different scaling regimes, corresponding to the
empirical evidence found for financial assets
\cite{iori_1,di_matteo_1,di_matteo_2}, can be identified within the
present model: one is an effective multiscaling regime, which is most
easily discussed in analytical terms for $t\leq M+1$, and the other is
an asymptotic Gaussian simple-scaling scenario, which prevails for
$t\gg M$ as a consequence of the central limit theorem mentioned above
\cite{mathpaper}.  We focus here on the former, which is directly
relevant for applications in finance.

The moment time-dependence of the aggregated return $X_1+\cdots+X_t$
is only ruled by the 
short-memory component if $t\leq M+1$, since
Eq.\ (\ref{ftsimple}) enables one to demonstrate \cite{supplementary}
that in such a case
\begin{eqnarray}
\nonumber m_q^X(t)&\equiv &\frac{\mathbb
  E[|X_1+\cdots+X_t|^q]}{\mathbb E[|X_1|^q]}\\ &=&\frac{\mathbb
  E\bigl[(a^2_{I_1}+\cdots+a^2_{I_t})^{\frac{q}{2}}\bigr]}{\mathbb
  E[a^q_{I_1}]}.
\label{eq_mqX}
\end{eqnarray}
Notice that the r.h.s.\ of Eq. (\ref{eq_mqX}) is well defined for any
$q$.  If the $X_t$'s PDF's are endowed with fat tails, this is not
true for the l.h.s.\ of the same equation. When $D<1/2$ and not too
small, effective scaling properties of the model follow from the fact
that $m_q^X$, although apparently a rather complex function of the
time, is well approximated by the power $t^{q\;H_q}$ for $t\le M+1$
\cite{supplementary}. The generalized Hurst-like exponent $H_q$ can be
computed using a least squares method over time.  Referring for
instance to a temporal window extending up to $t=31$ and adopting a
memory order $M\ge 30$, Fig.\ \ref{fig_multiscaling_model} displays
$H_q$ for different pairs of $D$ and $\nu$ values.  The exponent $H_q$
stays close to $1/2$ for low orders $q$ up to about $2/(1-2D)$,
denoting an initial simple-scaling regime. It recovers a dependence on
$q$ for larger moment orders, manifesting a multiscaling behavior. A
sharp result is found in the limit of small $\nu$, where $H_q=1/2$ for
$q\le 2/(1-2D)$ \cite{supplementary}.

In the perspective of a comparison of our model with data, a remark on
the scaling features of empirical financial data is in order (See also
\cite{peirano_1}).  While the simple-scaling behavior at low $q$ is a
stable and robust empirical evidence, multiscaling features occurring
at larger $q$ are sensibly dependent on the time series sample, for
series of a length comparable with that at our disposal for the
S\&P500 index. We report this observation in
Fig.\ \ref{fig_multiscaling_sep} with respect to the S\&P500 daily
time series. The empirical exponent $\overline H_q$ is here obtained
from the time-average estimation of $m_q^X$, as computed in the next
Section.  In turn, from the modeling point of view, the multiscaling
region in the moment order axis mostly overlaps the non-existing
moment region when fat-tailed distributions are involved.

\subsection{Volatility autocorrelation}
\label{sub_auto}

In view of financial applications, the volatility autocorrelation of
order $q$ can be introduced as the autocorrelation function of the
process $\{|X_t|^q\}_{t=1}^{\infty}$:
\begin{equation}
r_q^X(t)\equiv \frac{\mathbb E[|X_1|^q|X_t|^q]-\mathbb E[|X_1|^q]^2}{\mathbb E[|X_1|^{2q}]-\mathbb E[|X_1|^q]^2}.
\label{eq_rqX}
\end{equation}
Again, this autocorrelation is easily investigated for $t\le M+1$,
where the Markov-switching component alone determines its
decay. Indeed, thanks to the independence of the processes
$\{Y_t\}_{t=1}^{\infty}$ and $\{I_t\}_{t=1}^{\infty}$, $r_q^X(t)$ can
be rewritten as \cite{supplementary} $r_q^X(t)=u_q+v_q \;r_q^{a_I}(t)$
for $2\le t\le M+1$, while $r_q^X(1)=1$.  Here, $r_q^{a_I}$ is the
autocorrelation of $\{a_{I_t}^q\}_{t=1}^{\infty}$, and $u_q$, $v_q$
are two time-independent coefficients whose explicit expression is
provided in the Supplementary Material \cite{supplementary}.  We thus
see that the time dependence of $r_q^X$ comes from $r_q^{a_I}$ at
short time scales $t\leq M+1$. Interestingly, for $q$ small enough,
the smaller the restart probability $\nu$, the more correlations get
persistent: when $\nu$ approaches zero, we find $r_q^{a_I}(t)=1$ for
any $t$ if $q\le 1/(1-2D)$. Notice that this last threshold for the
moment order $q$ is now half of that previously discussed for the
simple-scaling behavior.

While the initial decay of the volatility autocorrelation $r_q^X$ is
strongly dependent on the parameter setting, in particular through the
ratio $u_q/v_q$, on time scales much larger than $M$, $r_q^X$ decays
exponentially fast, due to the strong mixing properties of our model
\cite{mathpaper}. In the Supplementary Material \cite{supplementary}
we show that this is indeed the case focusing on the function $\rho$
given by Eq.\ (\ref{rhoinv}) and the instance $q=2$, for which the
correlation decay rate can be explicitly computed. 

We conclude the Section with a remark concerning our
convention of referring to ``long-memory'' and ``short-memory''
processes, which contrasts with some common
use in the econometric literature. Indeed, within this literature a
process is said to possess long memory if the autocorrelation is not
summable in time \cite{Baillie}.
The asymptotic exponential decay of $r_1^Y(t)$ 
provided by our model \cite{supplementary} 
entails that the above sum is finite also for
the process $\{Y_t\}_{t=1}^{\infty}$.  
However, our convention stresses the different structure of the two
components.

\section{Model calibration}
\label{sec_calibration}

An important issue for the application of a model to time series
analysis is the implementation of efficient calibration protocols,
capable of identifying the model parameters which most effectively
reproduce a specific empirical evidence.  As anticipated, the
inclusion of both a long-memory and a short-memory part in our model
complicates the calibration procedure, because the two components
cannot be easily resolved along an empirical time series.  In order to
overcome this difficulty, we devise here a method based on the
comparison between empirical and theoretical moments, drawing on the
generalized method of moments \cite{alastair_1} and taking advantage
of the analytical structure of our model.

With the relatively limited amount of daily historical data available
for financial assets, the identification of the model parameters is
affected by large uncertainties.  Since our memory parameter $M$
establishes the time horizon over which the long-memory endogenous
dynamical dependence operates, we can choose to fix it on the basis of
the time scale associated with the specific application of interest.
Given $M$, we thus optimally exploit the simple analytical structure
within the time window $1\leq t\leq M$ for the calibration of the
remaining parameters. In order to present the procedure and to test
our model on real data, we refer here and in the following to the
advantageous function $\rho$ introduced in Section
\ref{sec_construction} by Eq.\ (\ref{rhoinv}). Once $M$ is fixed, the
further parameters to be estimated are four: the exponent $D>0$, the
restart probability $0<\nu\leq1$, and the parameters $\alpha>0$ and
$\beta>0$ identifying $\rho$.  For simplicity, we collect the first
three of them into the vector $\theta\equiv (D,\nu,\alpha)$ and we
denote by $\Theta$ its feasible range. The parameter $\beta$ plays a
minor role in the model since we only need it to fix the scale of
$X_t$'s fluctuations.

Given a time series $\{\overline x_t\}_{t=1}^T$ with empirical mean
zero, our calibration protocol is based on the idea of better
reproducing, within the model, its scaling and autocorrelation
features on times up to $M$. Thus, in a least square framework, we
choose those parameters which minimize the distance between the
theoretical $m_q^X(t)$ and $r_q^X(t)$, defined by Eqs.\ (\ref{eq_mqX})
and (\ref{eq_rqX}) respectively, and the corresponding empirical
estimations $\overline m_q^X(t)$ and $\overline r_q^X(t)$ in the
window $1\leq t\leq M$.  Such empirical estimations are obtained via
time averages over the available series. To illustrate the
computation, for instance we get $\overline m_q^X(t)$ as
$\mathcal{M}_q(t)/\mathcal{M}_q(1)$ with
\begin{equation}
\mathcal{M}_q(t)\equiv \frac{1}{T+1-t}\sum_{n=0}^{T-t}|\overline x_{n+1}+\cdots + \overline x_{n+t}|^q.
\end{equation}
We recall that the comparison between empirical time averages and
theoretical ensemble expectations is justified by the ergodicity of
our process \cite{mathpaper}.

Being properly normalized, $m_q^X$ and $r_q^X$ do not depend on the
scale parameter $\beta$. Denoting by $\mathcal{Q}$ the set of the
moment orders we consider for the calibration purposes, our parameter
estimation $\overline\theta\equiv (\overline
D,\overline\nu,\overline\alpha)$ results thus to be
\begin{eqnarray}
\nonumber
\overline \theta &=&\underset{\theta\in\Theta}{\arg\min}\biggl\{\biggr.
\sum_{q\in\mathcal{Q}}\sum_{t=1}^{M}\left[\frac{m_q^X(\theta;t)-\overline m_q^X(t)}{m_q^X(\theta;t)}\right]^2 +\\
&&~~~~~~~~~+\sum_{q\in\mathcal{Q}}\sum_{t=1}^{M}\left[\frac{r_q^X(\theta;t)-\overline r_q^X(t)}{r_q^X(\theta;t)}\right]^2\biggl.\biggr\},
\end{eqnarray}
where the dependence of $m_q^X$ and $r_q^X$ on $\theta$ is explicitly
indicated.  We have directly checked that this calibration procedure
precisely recovers the input parameters when applied to sufficiently
long time series simulated through the model.

\begin{figure}
\includegraphics[width=0.99\columnwidth]{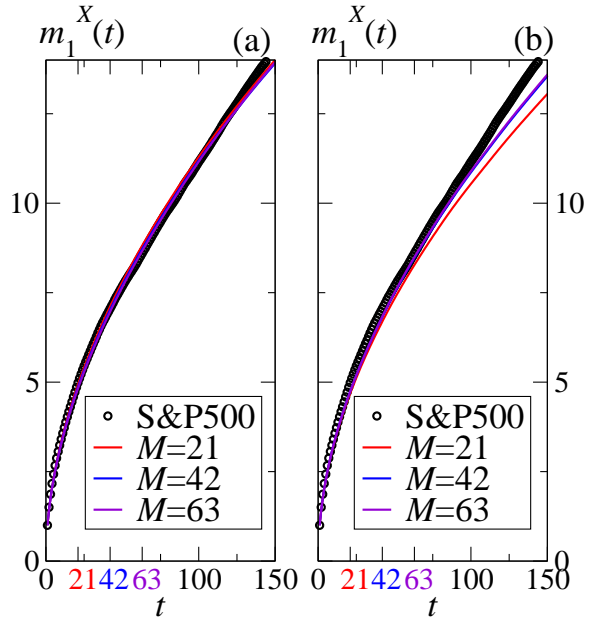}
\caption{ Calibration outcome in terms of the scaling indicator
  $m_1^X$ for different values of $M$, also reported in the abscissa
  with different colors. In (a) the model with $\rho$ given by
  Eq.\ (\ref{rhoinv}); in (b) the null model. 
}
\label{fig_moments}
\end{figure}

\begin{figure}
\includegraphics[width=0.99\columnwidth]{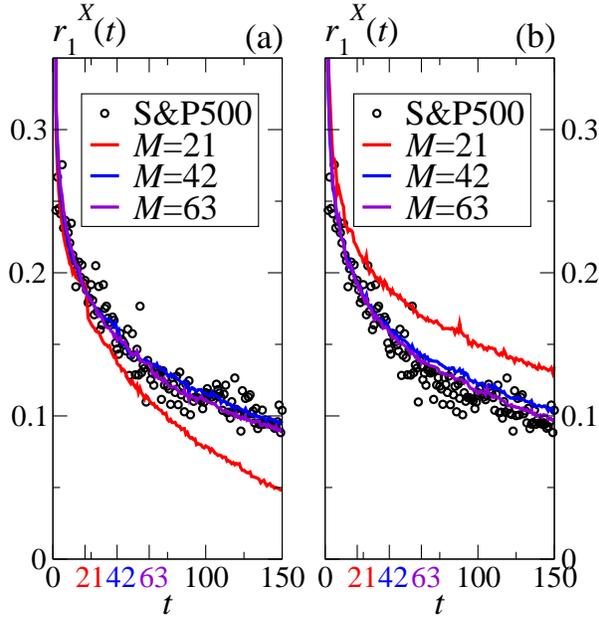}
\caption{ Calibration outcome in terms of the volatility
  autocorrelation $r_1^X$. In (a) the model with $\rho$ given by
  Eq.\ (\ref{rhoinv}); in (b) the null model.
}
\label{fig_autocorr}
\end{figure}

We close the calibration protocol providing a way of estimating the
parameter $\beta$. Once $\overline D$, $\overline\nu$, and
$\overline\alpha$ have been obtained, we can evaluate $\beta$ by
optimizing with respect to $e_q^X\equiv\mathbb E[|X_1|^q]$, where $D$,
$\nu$, and $\alpha$ are set equal to $\overline D$, $\overline\nu$,
and $\overline\alpha$ respectively.  Making explicit the dependence of
$e_q^X$ on $\beta$, the relationship $e_q^X(\beta)=e_q^X(1)\beta^q$ is
rather evident. If $\overline e_q^X$ denotes the empirical counterpart
of $e_q^X$, we then get our estimation $\overline\beta$ of $\beta$
through the formula
\begin{equation}
\overline \beta=\underset{\beta\in (0,\infty)}{\arg\min}\biggl\{
\sum_{q\in\mathcal{Q}}\left[\frac{e_q^X(\beta)-\overline e_q^X}{e_q^X(\beta)}\right]^2\biggr\}.
\end{equation}

Aiming at reducing the computational load of the parameter
estimations, in the present Paper we work out calibration with the
moment order $q=1$ only: $\mathcal{Q}=\{1\}$.
Fig.\ \ref{fig_moments}a and \ref{fig_autocorr}a report the result of
this protocol applied to the logarithmic increments of the daily
closures of S\&P500 from January 1st 1950 to December 31st 2010. We
set $\overline x_t\equiv\ln\overline s_t-\ln\overline s_{t-1}-\mu$ for
$t=1,\ldots,T$, being $T=15385$ and $\{\overline s_t\}_{t=0}^{T}$ the
considered S\&P500 time series.  The value of the drift $\mu$ is such
that $\sum_{t=1}^T\overline x_t=0$. In compliance with an application
we are developing to derivative pricing \cite{baldovin_4}, we have
chosen $M=21$ (the operating market days in one month) yielding
$(\overline
D,\overline\nu,\overline\alpha,\overline\beta)=(0.21,0.030,4.0,0.04)$,
$M=42$ (two months) giving $(\overline
D,\overline\nu,\overline\alpha,\overline\beta)=(0.19,0.011,4.5,0.07)$,
and $M=63$ (three months) for which $(\overline
D,\overline\nu,\overline\alpha,\overline\beta)=(0.16,0.004,5.5,0.14)$.
Notice how the calibrated model fits the S\&P500 scaling features and
the volatility autocorrelation well beyond $M$ in the case of two and
three months, whereas one month does not seem to be enough to get the
correct decay as soon as $t$ is larger than $21$.

\section{Comparison with S\&P500 index and null hypothesis}
\label{sec_main}

\begin{figure}
\includegraphics[width=0.99\columnwidth]{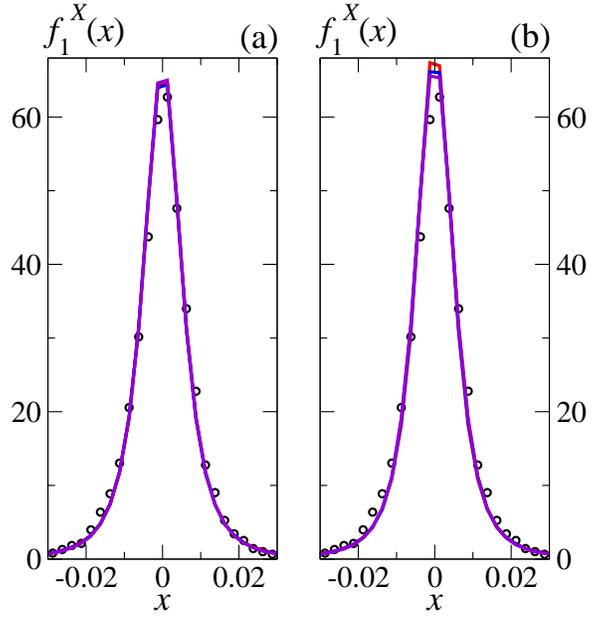}
\caption{
  Single-variable PDF comparison 
  between S\&P500 and calibrated complete model (a), and 
  between S\&P500 and the null model (b) in linear scale.
  Symbols and lines color code is as in the previous plots.
}
\label{fig_pdf}
\end{figure}

\begin{figure}
\includegraphics[width=0.99\columnwidth]{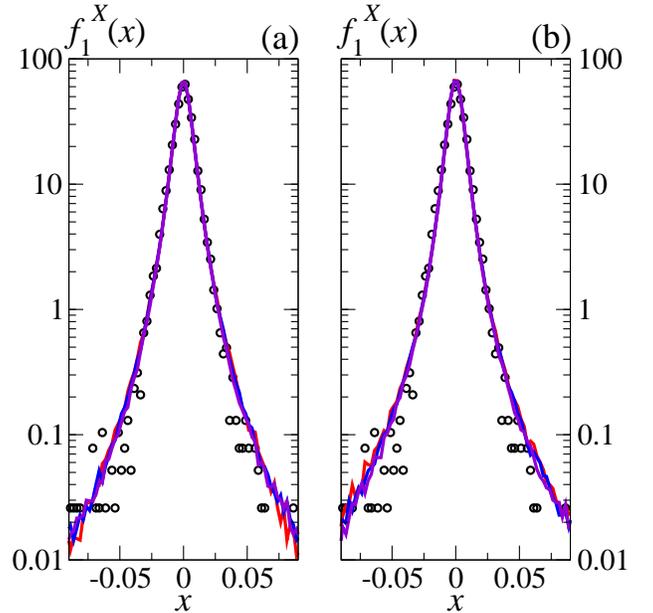}
\caption{
Single-variable PDF comparison between S\&P500 and calibrated complete model (a),
and between S\&P500 and the null model (b) in log scale.
Symbols and lines color code is as in the previous plots.  }
\label{fig_pdf_log}
\end{figure}

In order to put into context the performance of our model and to probe
the role of the memory $M$, here we consider, as the null hypothesis,
a limit version in which $\sigma$ is kept fixed to a constant value
$\sigma_0$ [$\rho(\sigma)=\delta(\sigma-\sigma_0)$], which turns out
to be the scale parameter.  From Eq.\ (\ref{gp_1}) we get that this
prescription replaces the auto-regressive component with a sequence of
independent normal variables, preventing the parameter $M$ from
playing any role
\cite{note_andreoli}.
Even if the null model
has no endogenous memory, for the sake of comparison we estimate its
parameters (first $D$ and $\nu$, and later $\sigma_0$) by means of the
procedure outlined in the previous Section and with the same values of
$M$ and $q$ used for the model characterized by the function $\rho$ of
Eq.\ (\ref{rhoinv}), which we name here ``the complete model''.
Figs.\ \ref{fig_moments}b and \ref{fig_autocorr}b show the outcome of
the calibration protocol, which gives the following results:
$(\overline D,\overline\nu,\overline\sigma_0)=(0.05,0.0001,1.01)$ for
$M=21$, $(\overline
D,\overline\nu,\overline\sigma_0)=(0.06,0.0002,0.62)$ for $M=42$, and
$(\overline D,\overline\nu,\overline\sigma_0)=(0.07,0.0003,0.45)$ when
$M=63$.  The figures indicate that calibration is slightly less
successful for the null model than for the complete one.

\begin{figure}
\includegraphics[width=0.99\columnwidth]{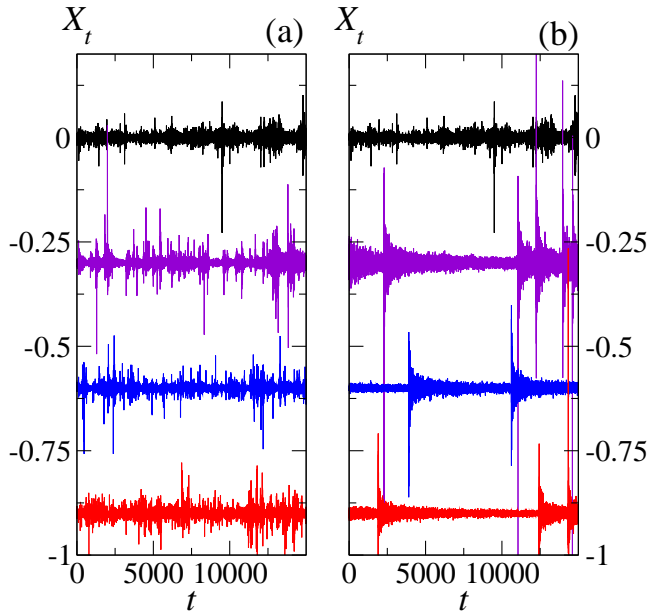}
\caption{
  Time series comparison between S\&P500 (black) and complete model (a), and 
  between S\&P500 (black) and the null model (b). To facilitate the inspection, 
  model time series are shifted by -0.3 ($M=63$), -0.6 ($M=42$), and -0.9 ($M=21$).
}
\label{fig_series}
\end{figure}

The unconditional return PDF of the S\&P500 is very well reproduced by
both the complete and the null calibrated models, both in the central
part and in the tails. We realize this fact by an inspection of the
linear and log plots of $f_1^X$ in Figs.\ \ref{fig_pdf} and
\ref{fig_pdf_log} respectively.  While the function $\rho$ defined by
Eq.\ (\ref{rhoinv}) endows the complete model with fat tails, setting
$\sigma=\sigma_0$ prevents the null model from recovering such a
feature from a strict mathematical standpoint. However, in Section
\ref{sub_PDF} we mentioned that with a small enough value of the
restart probability $\nu$ one recovers an effective fat tails scenario
when $D<1/2$. This circumstance explains why the null model reproduces
the empirical fat tails thanks to an estimated value of $\nu$ which is
one or two orders of magnitude smaller than the corresponding value
for the complete model.  The drawback is that a very small restart
probability entails very rare but high and strongly time-asymmetric
volatility bursts in the typical trajectories of the model, which are
not observed in the historical series.  Indeed,
Fig.\ \ref{fig_series}b, showing the comparison of typical simulated
realizations of the benchmark model with the S\&P500 time series,
reports the discrepancy between the S\&P500 and the null model paths,
where one can immediately identify the time restarts. In contrast,
once the auto-regressive component retains the memory of the previous
returns, the combined effect of more frequent restarts and of the
volatility clustering phenomenon produces typical trajectories which
are pretty similar to the historical S\&P500, as shown in
Fig.\ \ref{fig_series}a where the above comparison is proposed for the
complete model. Notice that restart events become here much harder to
identify.

\begin{figure}
\includegraphics[width=0.99\columnwidth]{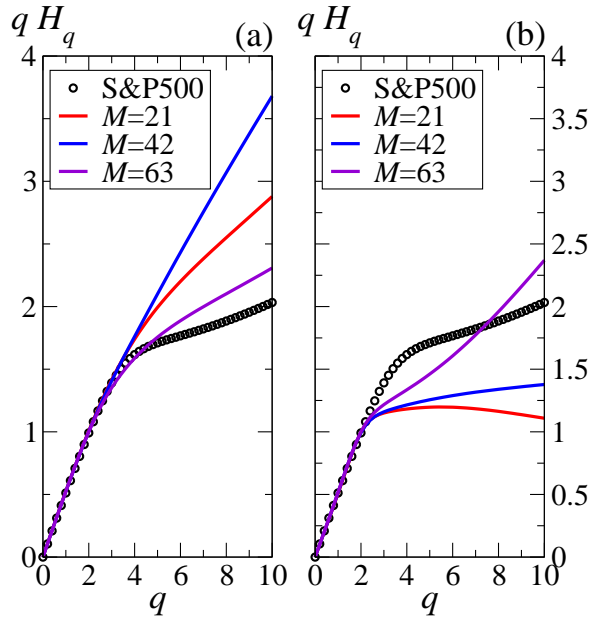}
\caption{Multiscaling comparison between the S\&P500 and the complete
  model (a), and between the S\&P500 and the null model (b).}
\label{fig_multiscaling}
\end{figure}

In order to recover the role of the exponent $D$, in
Figs.\ \ref{fig_multiscaling}a and \ref{fig_multiscaling}b we also
compare the aggregated return scaling features of the calibrated
models with those of the S\&P500 series. While, as anticipated in
Fig.\ \ref{fig_multiscaling_sep}, the empirical multiscaling regime is
very erratic and dependent on single extreme events, the
simple-scaling behavior ($\overline H_q\simeq1/2$) up to $q\simeq3$
seems a stable feature of the S\&P500. On the other hand, in Section
\ref{sub_scaling} we noticed that our model predicts that the latter
extends up to $q=2/(1-2D)$ at low values of $\nu$ when $D<1/2$,
irrespective of the function $\rho$. The complete model provides
$2/(1-2\overline D)=3.4$ for the calibration with $M=21$,
$2/(1-2\overline D)=3.2$ for $M=42$, and $2/(1-2\overline D)=2.9$ if
$M=63$, therefore showing a qualitative agreement with the empirical
evidence. The same cannot be said for the null model, which gives
$2/(1-2\overline D)$ close to 2 for all the three calibrations.

Financial time series are reported to break time-reversal invariance,
not only in terms of return-volatility correlation properties (e.g.,
the leverage effect \cite{bouchaud_1,bouchaud_lev}), but also in terms
of volatility-volatility correlations \cite{zumbach_2}.  Although in
the form discussed so far our model cannot explain the former (which
is an odd-even correlation), it can account for the latter since, as
we have already mentioned in Section \ref{sub_PDF}, the
Markov-switching component breaks the temporal symmetry through the
mechanism of time restarts. We thus conclude this Section considering
an even-even correlation, specifically the historical versus realized
volatility correlation
\cite{zumbach_2,borland_1,borland_2,zumbach_1,peirano_1}, and
assessing an asymmetry between the past and the future for the
calibrated complete model. In Section \ref{sec_improvements} we shall
discuss how to improve the present model in order to also take into
account the leverage effect.

\begin{figure*}
\includegraphics[width=0.9\columnwidth]{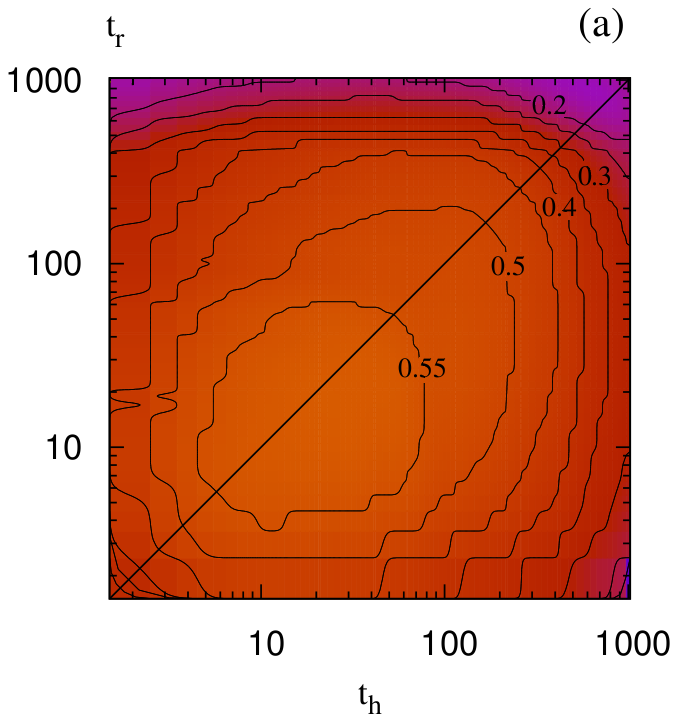}
\includegraphics[width=0.9\columnwidth]{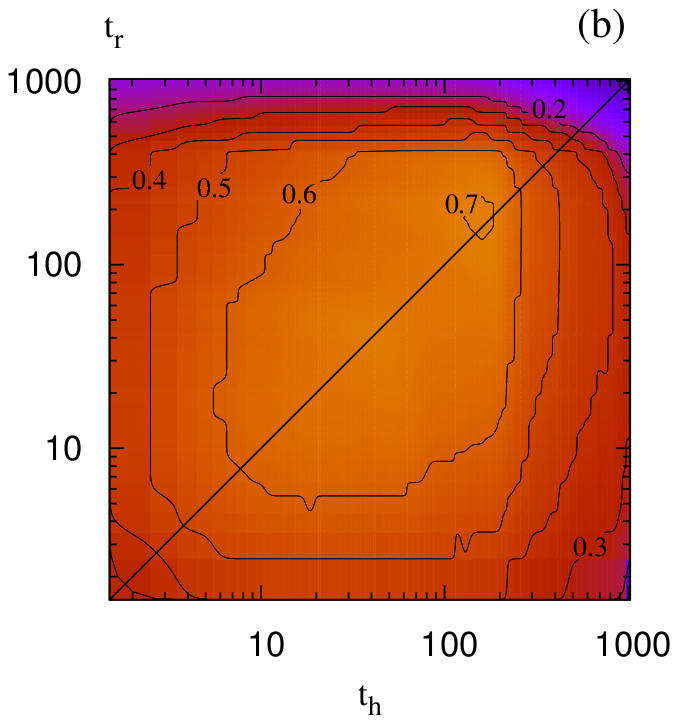} \\
\includegraphics[width=0.9\columnwidth]{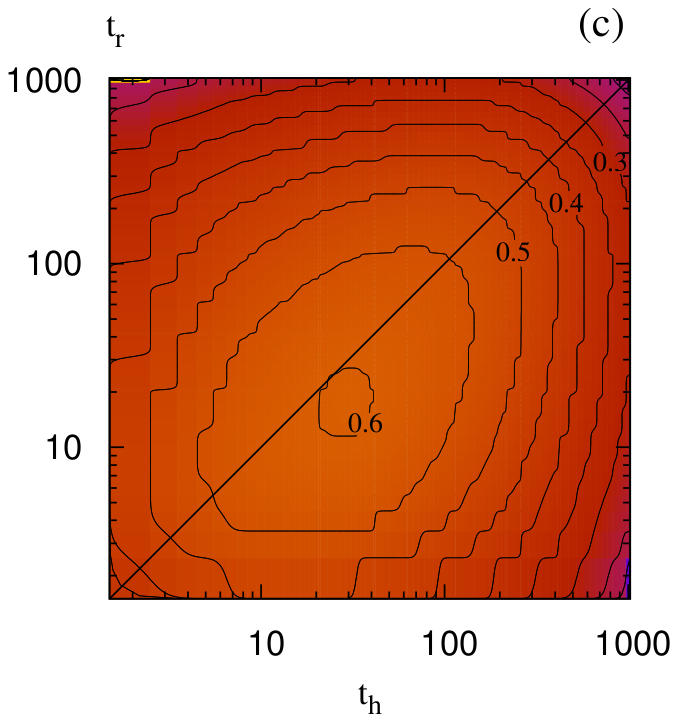}\\
\smallskip
\smallskip
\includegraphics[width=1.8\columnwidth]{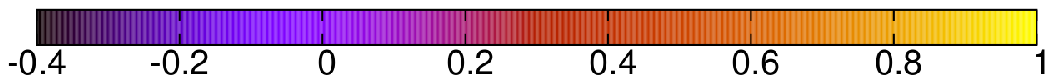}\\
\caption{
  Level curves plot of $\overline{\chi}(t_h,t_r)$, or volatility mug shots.
  (a) whole S\&P500 time series ($T=15385$);
  (b) second half of S\&P500 time series ($T=7694$); (c) model's
  prediction with the $M=42$-calibration. }
\label{fig_mug_shots}
\end{figure*}

Consider two consecutive time windows, named ``historical'' and 
``realized'', of width $t_h\ge 1$ and
$t_r\ge 1$, respectively. 
The associated ``historical volatility'' $S_{t_h}^h$ 
and ``realized volatility'' $S_{t_r}^r$ 
are defined as the random variables
\begin{equation}
S_{t_h}^h\equiv\sqrt{\frac{1}{t_h} \sum_{t=1}^{t_h} X_t^2 }
\end{equation}
and
\begin{equation}
S_{t_r}^r\equiv\sqrt{\frac{1}{t_r} \sum_{t=1}^{t_r} X_{t_h+t}^2 } \; .
\end{equation}
For a reversible process, 
the correlation between past and future volatilities \cite{lynch_1},
namely
\begin{equation}\label{eq_mug_shots}
\chi(t_h,t_r) \equiv \frac{\mathbb{E}\bigl[S_{t_h}^h S_{t_r}^r\bigr]-\mathbb{E}\bigl[S_{t_h}^h\bigr]\mathbb{E}\bigl[S_{t_r}^r\bigr]}
{\sqrt{{\rm var}\bigl[S_{t_h}^h\bigr]\;{\rm var}\bigl[S_{t_r}^r\bigr]}}\; ,
\end{equation}
is a symmetric function of the time horizons $t_h$ and $t_r$,
as one can easily verify starting from the
definition of reversibility given in Section \ref{sub_PDF}. In
contrast, the structure of its empirical estimation 
$\overline{\chi}(t_h,t_r)$ for the S\&P500 time
series shows some degrees of asymmetry.
This is highlighted by the level curves plot in 
Fig.\ \ref{fig_mug_shots}, also named ``volatility mug
shots'' \cite{borland_1,zumbach_1}.  
Such an asymmetry is however rather mild and sample
dependent, as illustrated by Figs.\ \ref{fig_mug_shots}a and
\ref{fig_mug_shots}b where the whole S\&P500 sample and the
second half only are exploited, respectively. As far as our model is concerned, at
variance with what pointed out in Ref.\ \cite{peirano_1} for a
different implementation of our ideas, we remark that such a mild time
asymmetry is consistently reproduced. 
For instance, Fig.\ \ref{fig_mug_shots}c displays the
level curves of $\chi(t_h,t_r)$ 
corresponding to the complete model  calibrated with $M=42$.

In the various comparisons outlined in this Section we have used
the average values defined by our calibrated model. 
In model-generated time series with a length of the order of that
of the available S\&P500 dataset, we have also inspected the 
fluctuations around these average values. 
In general, we have observed fluctuations that are consistent with
those associated to the sample-dependence of the S\&P500 time series.

\section{Long-memory and Short-memory volatility}
\label{sec_endex}

An interesting feature for a model of asset evolution is the
possibility of distinguishing between long-memory and short-memory
contributions to the volatility.  Since part of the short-memory
random effects may be attributed to the impact of external information
on the asset's time evolution, such a distinction is also related to
attempts in separating the endogenous and exogenous contributions to
the volatility
\cite{Cutler,Sornette_1,Sornette_2,Joulin,Filimonov_1,Hardiman,Filimonov_2}.
Indeed, although this should not be regarded as a clear cut
distinction, one may reasonably expect that long-memory contributions
could be ascribed to cooperative influences among the agents, whereas
random volatility switches may also come from news reaching the
market.  In our model, albeit intimately combined together, the
long-memory and short-memory components play their own distinct role
in reproducing realistic financial features. 
The question then naturally arises about the possibility of
identifying these two different contributions.  For this reason, we
propose here a procedure to localize the time restarts in a given
finite realization $\{\overline x_t\}_{t=1}^T$ of a process which is
assumed to be well represented by our model.  Once the restarts are
supposed to be known, we can identify the auto-regressive trajectory
$\{\overline y_t\}_{t=1}^T$, thus succeeding in distinguishing between
the two contributions.

\begin{figure}
\includegraphics[width=0.99\columnwidth]{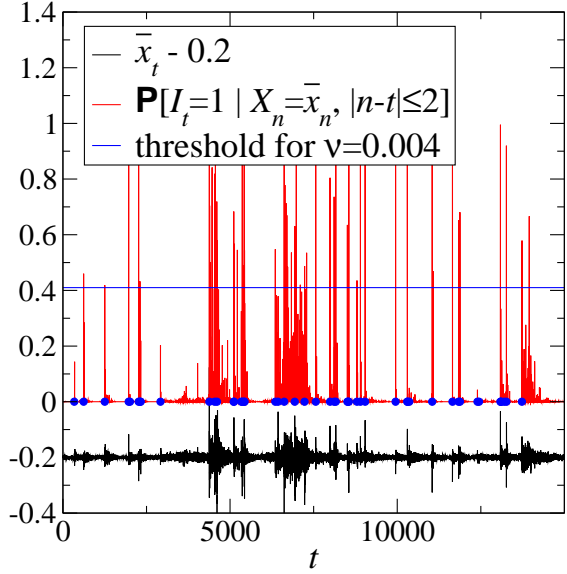}
\caption{ A procedure to locate the restart. In black is depicted 
  a $T=12000$-long time series generated by our model
  with the parameters $(\overline
  D,\overline\nu,\overline\alpha,\overline\beta)=(0.16,0.004,5.5,0.14)$,
  associated to the $M=63$ calibration. For convenience of inspection,
  the time series is vertically shifted by $-0.2$.  In red $\mathbb{P} [I_t=1 |
    X_n=\overline x_n,\;|n-t|\le2]$ is reported. The blue line
  corresponds to the threshold for which $\nu\,T$ restarts are
  detected; blue circles mark the true restarts generated by the dynamics.  }
\label{fig_restarts_model}
\end{figure}

\begin{figure}
\includegraphics[width=0.99\columnwidth]{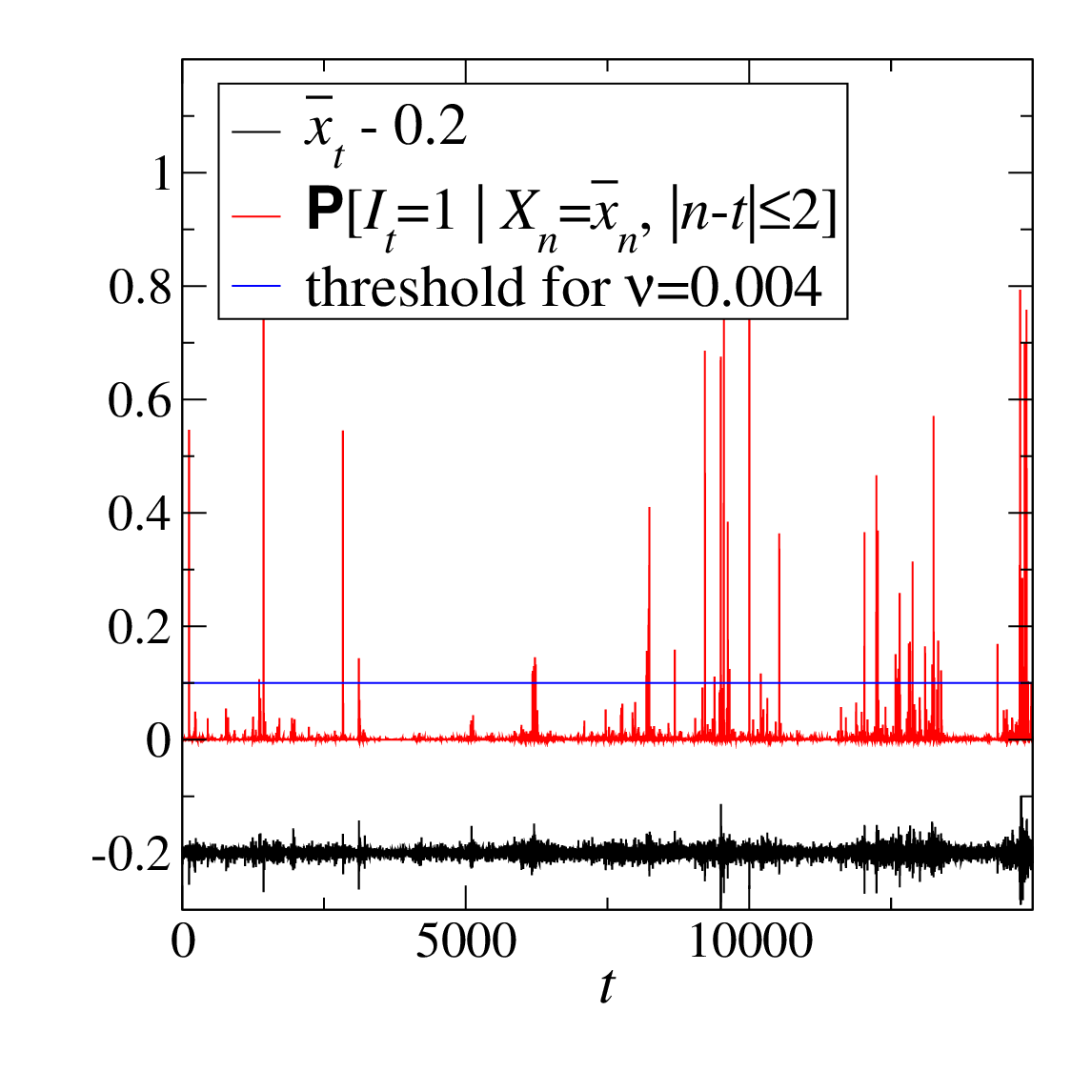}
\caption{
  Same as Fig.\ \ref{fig_restarts_model}, but applied to the S\&P500 historical
  time series. Here, of course, blue circles are absent.
}
\label{fig_restarts_sep}
\end{figure}

To the purpose of locating restarts, we consider the probability of
having a restart at a certain time $t$ conditioned to the information
available in a narrow time window centered in $t$ with half-width
$\tau$.  Namely,
\begin{equation} 
\mathbb{P} [I_t=1 | X_n=\overline x_n,\;|n-t|\le \tau].
\label{eq_ProbCondIt1}
\end{equation}
The time restarts can thus be tentatively associated to the peaks of
this probability. Since {\it a priori} we expect about $\nu T$
  time restarts, $T$ being the length of the considered time series,
  we associated them to the highest $\nu T$ peaks.  In
  Fig.\ \ref{fig_restarts_model} $\tau=2$ is used with respect to a
  model-generated time series. Conditioning the time restarts
identification to more time series values by taking a larger value of
$\tau$ would in principle provide better results.  In practice however
computational limitations force us to focus on small values of
$\tau$. Despite this restriction, 
in Fig.\ \ref{fig_restarts_model} we have been able to identify
exactly $60\%$ of the true restarts and about $70\%$ with an
uncertainty of two days.  Fig. \ref{fig_restarts_sep} displays the
result of the same ``time restarts analysis'' applied to the S\&P500
dataset.

\begin{figure}
\includegraphics[width=0.99\columnwidth]{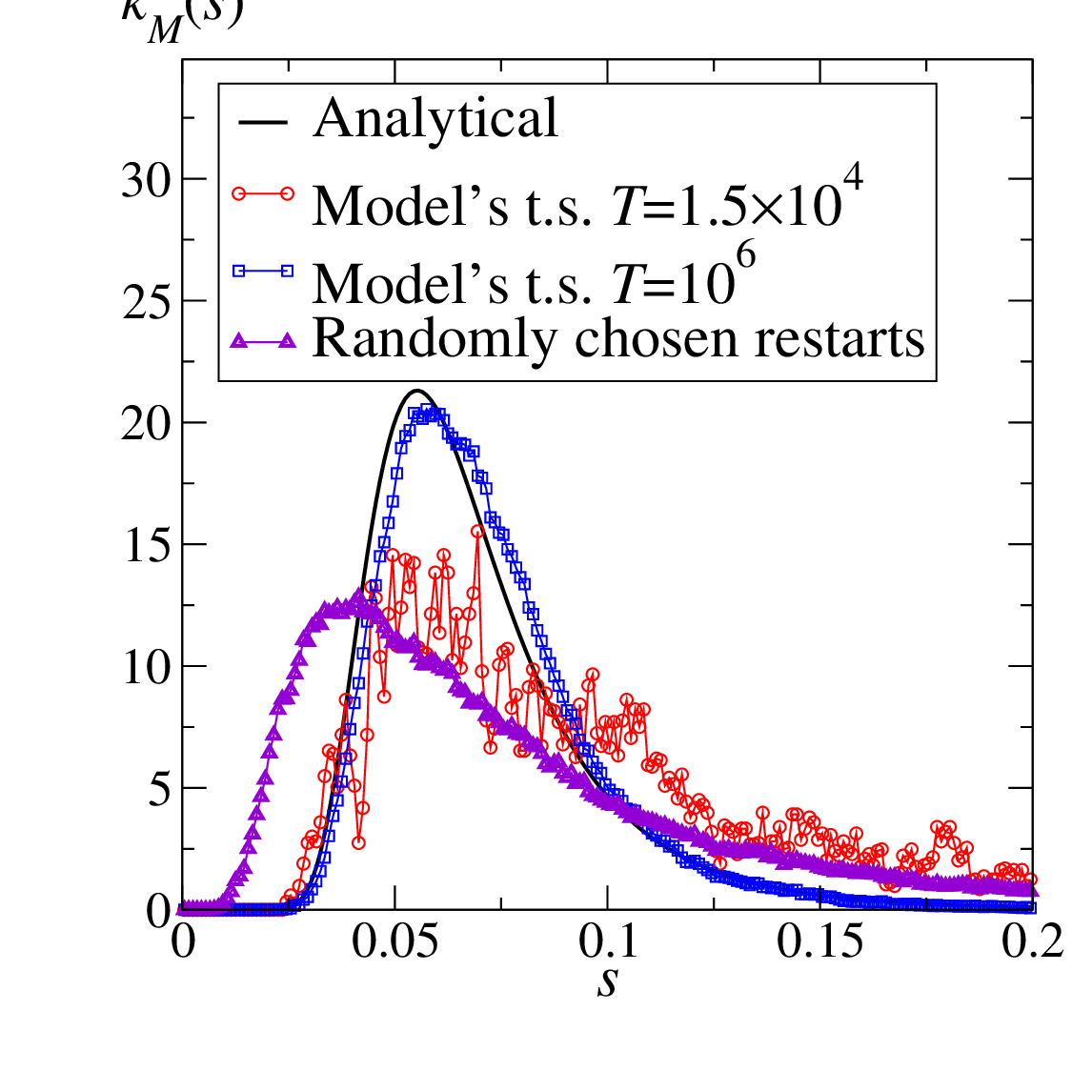}
\caption{Distribution of the long-memory component of the volatility:
  the continuous line is the theoretical 
  prediction, Eq.\ (\ref{eq_endVolTheo});
  circles and squares refer to two 
  time series of $T$ data generated by the model with the $M=63$ 
  calibration parameter set; triangles are obtained from the time series with 
  $T=10^6$, with the time restarts chosen randomly. 
}
\label{fig_EndogenousVolatility}
\end{figure}

\begin{figure}
\includegraphics[width=0.99\columnwidth]{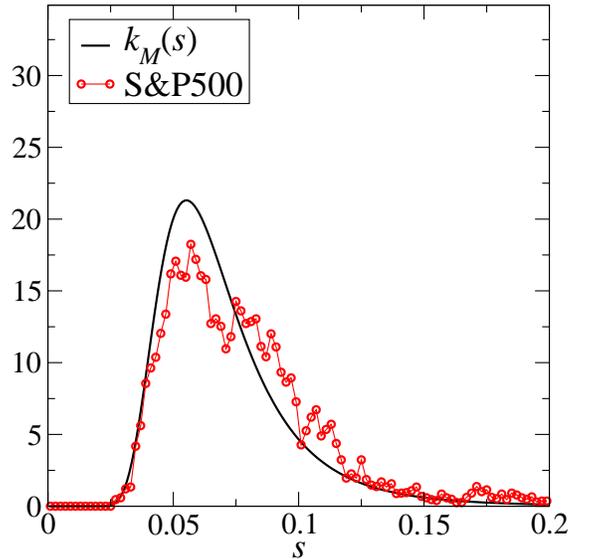}
\caption{Distribution of the long-memory component of the volatility for
  the S\&P500 dataset (circles), compared with the model-based prediction, 
  Eq.\ (\ref{eq_endVolTheo}), (full line). 
}
\label{fig_EndogenousVolatility_sep}
\end{figure}

Once the time restarts have been unveiled and the auto-regressive
trajectory $\{\overline y_t\}_{t=1}^T$ thus identified, we can analyze
the long-memory part of the volatility.  Within our model, a
convenient way of defining the long-memory volatility on the time
horizon $t$ is through the random variable
\begin{equation}
S_t \equiv  \sqrt{\frac{1}{t} \sum_{n=1}^t Y_n^2}.
\end{equation}
The PDF $k_t$ of this variable is easily obtained when $t \leq M+1$,
due to the fact that $f_t^Y$ reduces to a mixture of factorized
Gaussian densities with the same variance. It turns out to be
\begin{equation}
k_t(s)=\int_{0}^{\infty}d\sigma~\rho(\sigma)~\frac{2^{1-\frac{t}{2}} s^{t-1}t^{\frac{t}{2}}}{\Gamma\bigl(\frac{t}{2}\bigr)\sigma^t}~e^{-\frac{ts^2}{2\sigma^2}}.
\end{equation}
In particular, if the function $\rho$ is chosen according to
Eq.\ (\ref{rhoinv}), then $k_t$ is explicitly found as
\begin{equation}\label{eq_endVolTheo}
k_t(s) = 
\dfrac{2\beta^{\alpha} \, s^{t-1} \, t^{\frac{t}{2}} }
{B \left( \frac{\alpha}{2},\frac{t}{2}  \right) \,
 ( \beta^2 + s^2 t  )^{\frac{\alpha+t}{2}}},
\end{equation}
where $B$ is the Euler's Beta function. On the empirical side, the
distribution $k_t$ can be sampled from the estimated auto-regressive
path $\{\overline y_t\}_{t=1}^T$.
Fig.\ \ref{fig_EndogenousVolatility} shows a comparison between
theoretical and empirically detected long-memory component of the volatility
distributions for model-generated time series
with $t=M=63$.
Notice that as the
model's time series length $T$ increases, the outcome of the present
procedure becomes very close to the theoretical prediction in
Eq.\ ({\ref{eq_endVolTheo}}). 
This is particularly evident if, in place of 
using the
restarts  
obtained through Eq. (\ref{eq_ProbCondIt1}), 
we randomly choose them along the time series.
Finally, the consistency of the S\&P500
histogram with the theoretical predicition for $k_M$ 
(Fig.\ \ref{fig_EndogenousVolatility_sep})
points out that our procedure for
identifying the long-memory component of the volatility could be
successfully applied to the real market evolution, having sufficiently
long historical time series at disposal.

A distinction between long-memory and short-memory components of the
volatility is not a standard practice in finance. However, we think
that its consideration could open interesting perspectives in fields
like risk evaluation and market regulation.

\section{Improvements and further developments}
\label{sec_improvements}
Even if the present version of our model represents a significant
advancement in terms of
stylized-facts-reproduction-to-analytical-control ratio, some
important empirical features like the leverage effect and the skewness
of the return distribution are missing. Here we briefly discuss how
both these effects can be reproduced by suitable improvements of the
model.

The leverage effect refers to the presence in historical time series
of a negative odd-even correlation of the kind $\mathbb E[X_1 \,
  X_t^2]<0$. The model we have presented gives $\mathbb E[X_1 \,
  X_t^2]=0$ for any $t$ and, also, a symmetric returns
distribution. So far, in favor of an initial simplicity we have kept
the long-memory and short-memory components independent.  The
introduction of a dependence between these two processes, such as that
arising when the latter is assumed somehow affected by the past values
of the former (similarly, e.g., to the ideas outlined in
Ref.\ \cite{bouchaud_lev}) could produce non-zero sign-volatility
correlations like the leverage effect.  An appealing and potentially
interesting way of doing this within our mathematical construction may
simply consist in making time restarts dependent on the sign of the
auto-regressive endogenous component. We sketch some arguments about
this perspective.

Introducing the process $\{B_t\}_{t=1}^{\infty}$ of the signs of
$\{Y_t\}_{t=1}^{\infty}$, defined as $B_t=1$ if $Y_t\ge 0$ and
$B_t=-1$ if $Y_t<0$, in the Supplementary Material
\cite{supplementary} we show that $\{B_t\}_{t=1}^{\infty}$ and the
sequence $\{|Y_t|\}_{t=1}^{\infty}$ of the magnitude of $Y_t$'s are
mutually independent for the model considered so far. Moreover
$\{B_t\}_{t=1}^{\infty}$ results in a sequence of i.i.d. binary
variables with $\mathbb{P}[B_1=1]=1/2$, telling us that we have been
tossing a fair coin to decide the sign of returns. These
considerations allow one to recast our model as $X_t=a_{I_t}B_t|Y_t|$
with $\{I_t\}_{t=1}^{\infty}$, $\{B_t\}_{t=1}^{\infty}$, and
$\{Y_t\}_{t=1}^{\infty}$ independent from each other.

In order to improve the model, we could then think in general at
different alternatives. We could assume that the $B_t$'s take vale
different from $-1$ and $+1$ as, e.g., in Ref.\ \cite{eisler_1}.
Also, we could draw $B_{t+1}$ independently of the
past events, but making the restart occurrence $I_{t+1}=1$ dependent
on the value of $B_t$. In such a case, the process
$\{(I_t,B_t)\}_{t=1}^{\infty}$ would result in a bivariate Markov
chain, still independent of $\{Y_t\}_{t=1}^{\infty}$.  We already know
that a simple setting of this kind guarantees the martingale
character, the stationarity, and the mixing properties of the returns'
process $\{X_t\}_{t=1}^{\infty}$ defined as $X_t\equiv
a_{I_t}B_t|Y_t|$. At the same time, a skewness in the return
distribution is recovered by properly choosing the values assumed by
the $B_t$'s. The leverage effect occurs then making negative returns
more likely followed by a time restart than positive ones.  Work is in
progress along these lines.

Coming back to the model discussed in the present Paper, an
interesting applicative perspective is the fact that its analytical
handiness permits the derivation of closed-formulas for derivative
pricing and the associated hedging strategy.  As pointed out in
\cite{peirano_1}, in the presence of a Gaussian mixture process for
the underlying asset an obvious way of obtaining an arbitrage-free
option price is by taking the average Black-Scholes price
\cite{black_1,bouchaud_1,hull_1} according to the variance measure of
the mixture.  In the present approach, such a basic idea must be
shaped in order to take into account two basic facts. In first place,
the auto-regressive endogenous component implies that an effective
variance measure of the Gaussian mixture is conditioned by the
previous endogenous values of the process.  On the other side, the
Markov-switching process strongly influences the volatility.  Thus, an
effective way of identifying time restarts according to the scheme
discussed in Section \ref{sec_endex} must be developed.  In a related
work in progress \cite{baldovin_6} we have been able to successfully
tackle these two aspects and to produce an equivalent martingale
measure which allows one to derive European option prices
\cite{hull_1} in a closed form and to associate a natural hedging
strategy with the underlying asset dynamics.

\section{Conclusions}
\label{sec_conclusions}
Scaling and long range dependence have played a major role in the
recent development of stochastic models of financial assets dynamics.
This development proceeded parallel to the progressive realization
that indeed scaling and multiscaling properties are themselves
relevant stylized facts.
A key achievement has been the multifractal model of asset returns
(MMAR) proposed by Mandelbrot and coworkers \cite{mandelbrot_1}. This
model introduced important features, like the possibility of
multiscaling return distributions with finite variance and the long
range dependence in the volatility, with uncorrelated returns. This
long range dependence had been previously a peculiarity of ARCH or GARCH
type models \cite{engle_1,bollerslev_2,bollerslev_1,tsay_1}, widely
used in empirical finance.  The difficulties mainly arising from the
strict time reversal invariance of the MMAR has been overcome by
subsequent proposals of multi-time-scale models
\cite{zumbach_1,borland_2} which are somehow intermediate between
GARCH processes and descriptions based on multiplicative cascades.
However, a limitation of all the approaches mentioned above is due to
the scarce analytical tractability and the difficulty in efficiently
expressing the conditioning effect of past histories when applying
them to VaR estimates or option pricing.

The model we presented here addresses the problem first posed by
Bachelier over a century ago \cite{bachelier_1} and opens some
interesting perspectives.  From a methodological point of view, due to the
roots in renormalization theory, it offers an example where scaling
becomes a guiding criterion for the construction of a meaningful
dynamics.  This direction appears quite natural if we look at the
development of complex systems theory in statistical physics. Scaling
is normally regarded as a tool for unconditioned forecasting. Thanks
to our renormalization group philosophy, here we have shown that
scaling can also be exploited in order to obtain conditioned forecasting,
which is of major importance in finance. This conditioned forecasting
potential is based on the multivariate price return distributions like
Eqs.\ (\ref{ft}) and (\ref{ftsimple}), which one can construct 
on the basis of scaling properties.

The coexistence of exogenous and endogenous effects driving the
dynamics of the markets has been recognized since long.  Indeed, the
variations of the assets' price and volatility cannot be explained
only on the basis of arrival of new information on the market. A
remarkable feature of our model is the fact that it embodies a natural
and sound distinction between the long-memory endogenous influences
and the short memory, partially exogenous ones on the volatility. Even
if the distinction is model-based, the comparison with the S\&P500
dataset has shown consistency with historical data.

In the relatively simple form discussed in this Paper, our model has
important requisites for opening the way to useful applications. One
of these applications, namely a closed-form formulation for pricing
derivative assets, is presently under development
\cite{baldovin_6}. Indeed, in view of the capability to account for a
considerable number of stylized facts, our model maintains a high degree of
mathematical tractability. This tractability allows to rigorously
derive important mathematical properties of the process and to set up
successful calibration procedures.

A deep connection of our approach with ARCH models
\cite{engle_1,bollerslev_2,bollerslev_1,tsay_1} is the fact that we
identify an auto-regressive scheme as a natural one on which to base
the ergodic and stationary dynamics of our endogenous
process. Remarkably, we are naturally led to this choice 
following our criteria based on scaling and on the quest for
ergodicity and stationarity. 

Our modeling is not based on a ``microscopic'', agent based
description \cite{lux_1,lebaron_1,alfi_1}, which should be regarded as
a most fundamental and advanced stage at which to test the potential
of statistical physics methods in finance. However, we believe that
our results open in the field novel perspectives thanks to the
application of one of the most powerful methods available so far for
the study of complexity in physics, the renormalization group
approach.
This approach provides an original, valuable insight into the 
statistical texture of return fluctuations, which is a key requisite
for successful stochastic modeling.

\section*{Acknowledgments}
We would like to thank J.-P. Bouchaud and M. Caporin for 
useful discussions. 
This work is supported by 
``Fondazione Cassa di Risparmio di Padova e Rovigo'' within the 
2008-2009 ``Progetti di Eccellenza'' program.

\onecolumngrid
\medskip
\newpage
\setcounter{equation}{0}
\setcounter{section}{0}
\setcounter{figure}{0}

\begin{center}
{\bf \large Scaling symmetry, renormalization, and time series modeling \\ Supplementary Material}
\\
\medskip
Marco Zamparo \\
{\it \small Dipartimento di Fisica, Sezione INFN, CNISM, and Universit\`a di Padova, Via Marzolo 8, I-35131 Padova, Italy and \\ HuGeF, Via Nizza 52, 10126 Torino, Italy} \\
\medskip
Fulvio Baldovin, Michele Caraglio, and Attilio L. Stella \\
{\it \small Dipartimento di Fisica, Sezione INFN, CNISM, and Universit\`a di Padova, Via Marzolo 8, I-35131 Padova, Italy}
\setcounter{page}{1}
\end{center}

\section{Introduction}
This Supplementary Material provides proofs of the model properties
announced in the Main Text. In Section \ref{sec:pdf} the joint PDF's
of the process $\X$ are derived from the definition of the long-memory
and short-memory components $\Y$ and $\I$, respectively. Section
\ref{sec:stationarity} is devoted to prove the stationarity of $\X$,
while in Section \ref{sec:irreversibility} we show that $\Y$ is a
reversible sequence and that the series of its sign is independent of
$\{|Y_t|\}_{t=1}^{\infty}$.  In Sections \ref{sec:tail} and
\ref{sec:scaling} we reconsider the issue of the tail of the
single-variable PDF $f_1^X$ and the scaling features of our model
supplying a more complete study.  In Section \ref{sec:scaling} we
prove some properties of what we called the ``null'' and the
``complete'' models in the Main Text, in particular showing that the
endogenous component $\Y$ of the complete model [with the choice in
  Eq. (20) of the Main Text for the function $\rho$] is an ARCH
process. Lastly, Section \ref{sec:correlation} contains a detailed
analysis of the autocorrelation $r_q^X$.

\section{PDF's associated with the model}
\label{sec:pdf}
Here we derive the joint PDF's of the increments $\X$,
reported in Eq.\ (\PDFpaper) in the Main Text.
The derivation exploits the features
of the hidden processes $\Y$ and $\I$. To this and further purposes,
we shall work with the expectation values of test functions.  Unless
explicitly stated, we will implicitly assume that such expectation
values exist.

Given $t\ge 1$ and a test function $F$ on $\Real^t$, we have
\begin{eqnarray}
\Att[F(X_{1},\ldots,X_{t})] 
&=&\Att\bigl[F\bigl(a_{I_{1}}Y_{1},\ldots,a_{I_{t}}Y_t\bigr)\bigr]
\nonumber\\
&=&\sum_{i_{1}=1}^\infty\cdots\sum_{i_{t}=1}^\infty
\Att\bigl[F\bigl(a_{i_1}Y_{1},\ldots,a_{i_t}Y_{t}\bigr)~\delta_{I_{1}i_1}\cdots\delta_{I_{t}i_t}\bigr]~,
\label{PDFprima}
\end{eqnarray}
where as usual $\delta$ denotes Kronecker's symbol.  On the other
hand, recalling that $\Y$ and $\I$ are mutually independent, the
following factorization holds:
\begin{eqnarray}
\Att\bigl[F\bigl(a_{i_1}Y_{1},\ldots,a_{i_t}Y_{t}\bigr)~\delta_{I_{1}i_1}\cdots\delta_{I_{t}i_t}\bigr]
&=&\Att\bigl[F(a_{i_1}Y_{1},\ldots,a_{i_t}Y_{t}\bigr)]\cdot\Att[\delta_{I_{1}i_1}\cdots\delta_{I_{t}i_t}]
\nonumber\\
&=&\Att\bigl[F\bigl(a_{i_1}Y_{1},\ldots,a_{i_t}Y_{t}\bigr)]\cdot\Prob[I_{1}=i_1,\cdots,I_{t}=i_t]~.
\end{eqnarray}
Thus, plugging this in Eq.\ (\ref{PDFprima}), we obtain the chain of
identities
\begin{eqnarray}
\nonumber
\Att[F(X_{1},\ldots,X_{t})]
&=&\sum_{i_{1}=1}^\infty\cdots\sum_{i_{t}=1}^\infty\Att\bigl[F\bigl(a_{i_1}Y_{1},\ldots,a_{i_t}Y_{t})]
\cdot\Prob[I_{1}=i_1,\cdots,I_{t}=i_t]\\
\nonumber
&=&\sum_{i_{1}=1}^\infty\cdots\sum_{i_{t}=1}^\infty \Prob[I_{1}=i_1,\cdots,I_{t}=i_t]\cdot\\
\nonumber
&\cdot&\int_{\Real}d y_1\cdots\int_{\Real}d y_t~
F\bigl(a_{i_1}y_{1},\ldots,a_{i_t}y_{t}\bigr)~f_t^Y(y_1,\ldots,y_t)\\
\nonumber
&=&\int_{\Real}dx_1\cdots\int_{\Real}dx_t~F(x_{1},\ldots,x_{t})\cdot\\
&\cdot&\sum_{i_{1}=1}^\infty\cdots\sum_{i_{t}=1}^\infty \Prob[I_{1}=i_1,\cdots,I_{t}=i_t]~
\frac{f_t^Y\bigl(x_1/a_{i_1},\ldots,x_t/a_{i_t}\bigr)}{a_{i_1}\cdots a_{i_t}}~.
\end{eqnarray}
The last equality is the consequence of a simple change of
variables in the integrals. This result, combined with the arbitrariness
of $F$, clearly shows that the joint probability density distribution
of $(X_1,\ldots,X_t)$ is
\begin{eqnarray}
\nonumber
f_t^X(x_1,\ldots,x_t)
\nonumber
&\equiv&\sum_{i_{1}=1}^\infty\cdots\sum_{i_{t}=1}^\infty \Prob[I_{1}=i_1,\cdots,I_{t}=i_t]~
\frac{f_t^Y\bigl(x_1/a_{i_1},\ldots,x_t/a(i_t)\bigr)}{a_{i_1}\cdots a_{i_t}}\\
&=&\sum_{i_{1}=1}^\infty\cdots\sum_{i_{t}=1}^\infty 
W(i_t,i_{t-1})\cdots W(i_2,i_1)\pi(i_1)~\frac{f_t^Y\bigl(x_1/a_{i_1},\ldots,x_t/a_{i_t}\bigr)}{a_{i_1}\cdots a_{i_t}}~.
\label{ft}
\end{eqnarray}

As far as the density $f_t^Y$ is concerned, let us recall that
$f_t^Y=\varphi_t$ for $t\le M+1$ where
\begin{equation}
\varphi_t(y_1,\ldots,y_t)=\int_{0}^{\infty}d\sigma~\rho(\sigma)~\prod_{n=1}^t\mathcal N_{\sigma}(y_n)~.
\label{phiY}
\end{equation}
Moreover, solving Eq.\ (\EQPDFYpaper) of the Main Text, we explicitly get
$f_t^Y$ also when $t>M+1$ as
\begin{equation}
f_t^Y(y_1,\ldots,y_{t})=\frac{\displaystyle{\prod_{n=1}^{t-M}\varphi_{M+1}(y_n,\ldots,y_{n+M})}}
{\displaystyle{\prod_{n=2}^{t-M}\varphi_M(y_n,\ldots,y_{n+M-1})}}~.
\label{gesplic}
\end{equation}

\section{Stationarity}
\label{sec:stationarity}
This Section is devoted to prove the strict stationarity of
the process $\X$, for which we must verify that
$(X_n,\ldots,X_{n+t-1})$ is distributed as $(X_1,\ldots,X_t)$ for any
$n\ge 1$ and $t\ge 1$.  As a matter of fact, $\X$ inherits this
property from the hidden processes $\Y$ and $\I$ and so proving the
strict stationary of $\Y$ and $\I$ is the main issue. Let us
assume for a moment that we know that $(Y_n,\ldots,Y_{n+t-1})$ and
$(I_n,\ldots,I_{n+t-1})$ are distributed as $(Y_1,\ldots,Y_{t})$ and
$(I_1,\ldots,I_{t})$, respectively, for any $n\ge 1$ and $t\ge 1$. Then,
given $n$, $t$, and a test function $F$ on $\Real^{t}$ and exploiting
again the independence between $\Y$ and $\I$, we obtain
\begin{eqnarray}
\nonumber
\Att[F(X_n,\ldots,X_{n+t-1})]&=&
\Att\bigl[F\bigl(a_{I_n}Y_n,\ldots,a_{I_{n+t-1}}Y_{n+t-1}\bigr)\bigr]\\
\nonumber
&=&\sum_{i_{1}=1}^\infty\cdots\sum_{i_{t}=1}^\infty\Att\bigl[F\bigl(a_{i_1}Y_n,\ldots,a_{i_{t}}Y_{n+t-1}\bigr)\bigr]
\cdot\Prob[I_n=i_1,\ldots,I_{n+t-1}=i_{t}]\\
\nonumber
&=&\sum_{i_{1}=1}^\infty\cdots\sum_{i_{t}=1}^\infty
\Att\bigl[F\bigl(a_{i_1}Y_1,\ldots,a_{i_{t}}Y_{t}\bigr)\bigr]\cdot\Prob[I_1=i_1,\cdots,I_{t}=i_{t}]\\
&=&\Att[F(X_1,\ldots,X_{t})]~,
\label{stepX}
\end{eqnarray}
where the third equality is due to the hypothesis of stationarity of
both $\Y$ and $\I$.  The arbitrariness of $F$ then tells us that
$(X_n,\ldots,X_{n+t-1})$ is distributed as $(X_1,\ldots,X_{t})$. 

The stationarity of $\I$ was already discussed in the Main Text and is the
consequence of the fact that $\pi$ is the invariant distribution of
$W$. Thus, now we only have to analyze the process $\Y$.  In order to
prove the strict stationarity of this process, let us observe that for
any $n>M+1$, isolating the first terms in the products of
Eq.\ (\ref{gesplic}), we get the identity
\begin{equation}
f_n^Y(y_1,\ldots,y_{n})=\frac{\varphi_{M+1}(y_1,\ldots,y_{M+1})}
{\varphi_M(y_2,\ldots,y_{M+1})}f_{n-1}^Y(y_2,\ldots,y_n)~.
\end{equation}
Then, the fact that
\begin{equation}
\int_{\Real}dy_1~\varphi_{M+1}(y_1,\ldots,y_{M+1})=\varphi_M(y_2,\ldots,y_{M+1})
\end{equation}
leads us to the result
\begin{equation}
\int_{\Real}dy_1~f^Y_n(y_1,\ldots,y_n)=f^Y_{n-1}(y_2,\ldots,y_n)~,
\label{eq_reduction}
\end{equation}
which is also valid for $n\leq M+1$, where $f_n^Y=\varphi_n$, and
hence for any $n$. This relation allows us to prove that
\begin{equation}
\Att[F(Y_{n+1},\ldots,Y_{n+t})]=\Att[F(Y_{n},\ldots,Y_{n+t-1})]
\label{statY1}
\end{equation}
for any $n\ge 1$ and any function $F$ on $\Real^{t}$. Indeed
\begin{eqnarray}
\nonumber
\Att[F(Y_{n+1},\ldots,Y_{n+t})]&=&\int_{\Real}d y_1\cdots \int_{\Real}d y_{n+t}~F(y_{n+1},\ldots,y_{n+t})~
f^Y_{n+t}(y_1,\ldots,y_{n+t})\\
\nonumber
&=&\int_{\Real}dy_2\cdots \int_{\Real} dy_{n+t}~F(y_{n+1},\ldots,y_{n+t})~f^Y_{n+t-1}(y_2,\ldots,y_{n+t})\\
\nonumber
&=& \int_{\Real}d y_{1}\cdots \int_{\Real}dy_{n+t-1}~F(y_{n},\ldots,y_{n+t-1})~f^Y_{n+t-1}(y_{1},\ldots,y_{n+t-1})\\
&=&\Att[F(Y_{n},\ldots,Y_{n+t-1})]~,
\end{eqnarray}
where we have made use of Eq.\ (\ref{eq_reduction}) to obtain the
second equality and we have just re--labeled the variables to get
the third. The iteration of Eq.\ (\ref{statY1}) then provides
\begin{equation}
\Att[F(Y_{n},\ldots,Y_{n+t-1})]=\Att[F(Y_{1},\ldots,Y_{t})]~,
\end{equation}
which states the stationarity of the process $\{Y_t\}_{t=1}^{\infty}$.
$\qquad\Box$

\section{Reversibility and sign-magnitude independence of the endogenous component}
\label{sec:irreversibility}
In the Main Text we pointed out that the process $\Y$ is reversible,
namely that $(Y_t,Y_{t-1},\ldots,Y_1)$ is distributed as
$(Y_1,\ldots,Y_{t-1},Y_t)$ for any $t\ge 1$. Here we provide the proof
verifying that
\begin{equation}
f_t^Y(y_t,y_{t-1},\ldots,y_1)=f^Y_t(y_1,\ldots,y_{t-1},y_t)
\label{frev}
\end{equation}
for any $t\ge 1$ and $(y_1,\ldots,y_{t-1},y_t)\in\Real^t$.  This
identity descends from the invariance of $\varphi_t$ with respect to
permutations of its arguments and is evident if $t\leq M+1$. At the
same time, when $t>M+1$, replacing $(y_1,\ldots,y_{t-1},y_t)$ with
$(y_t,y_{t-1},\ldots,y_1)$ in Eq.\ (\ref{gesplic}) and rearranging the
indexes, we obtain
\begin{eqnarray}
\nonumber
f^Y_t(y_t,y_{t-1},\ldots,y_{1})&=&\frac{\displaystyle{\prod_{n=1}^{t-M}\varphi_{M+1}(y_{t-n+1},\ldots,y_{t-n-M+1})}}
{\displaystyle{\prod_{n=2}^{t-M}\varphi_M(y_{t-n+1},\ldots,y_{t-n-M+2})}}\\
&=&\frac{\displaystyle{\prod_{n=1}^{t-M}\varphi_{M+1}(y_{n+M},\ldots,y_{n})}}
{\displaystyle{\prod_{n=2}^{t-M}\varphi_M(y_{n+M-1},\ldots,y_{n})}}~.
\end{eqnarray}
The exchangeability of the arguments of $\varphi_{M+1}$ and $\varphi_M$, again, gives
Eq.\ (\ref{frev}).
$\qquad\Box$

In Section VIII of the Main Text, in order to propose possible extensions of the model,  
we also mentioned that the sign and the magnitude of
$\Y$ constitute two independent processes, the former being a sequence
of i.i.d. binary variables taking values in
$\mathbb{Z}_2\equiv\{-1,+1\}$ with equal probabilities. This fact
follows from the symmetry of $f_t^Y(y_1,\ldots,y_t)$ with respect to
any of its arguments.  
Setting $B_t=\text{sgn}(Y_t)$ with
$\text{sgn}(y)=1$ if $y\ge 0$ and $\text{sgn}(y)=-1$ if $y<0$, to
prove the above two statements 
we have to check that for any $t\ge 1$ and any test functions
$F$ on $\mathbb{Z}_2^t$ and $G$ on $\Real^t$ the identity
\begin{equation}
\Att[F(B_1,\ldots,B_t)~G(|Y_1|,\ldots,|Y_t|)]=\Att[F(B_1,\ldots,B_t)]\cdot\Att[G(|Y_1|,\ldots,|Y_t|)]
\label{eqB|Y|}
\end{equation}
and the relation
\begin{equation}
\Att[F(B_1,\ldots,B_t)]=2^{-t}\sum_{b_1\in\mathbb{Z}_2}\cdots\sum_{b_t\in\mathbb{Z}_2} F(b_1,\ldots,b_t)
\label{eqBdist}
\end{equation}
hold. 

For a general $F$ on $\Real^t$ we have the simple equality
\begin{equation}
\int_{\Real}dy_1\cdots\int_{\Real}dy_t~F(y_1,\ldots,y_t)=\int_0^{\infty}dy_1\cdots\int_0^{\infty}dy_t
\sum_{b_1\in\mathbb{Z}_2}\cdots\sum_{b_t\in\mathbb{Z}_2} F(b_1y_1,\ldots,b_ty_t)~.
\label{eqFb}
\end{equation}
Thus, we find that
\begin{eqnarray}
\nonumber
\Att[F(B_1,\ldots,B_t)~G(|Y_1|,\ldots,|Y_t|)]&=&\int_{\Real}dy_1\cdots\int_{\Real}dy_t~
F(\text{sgn}(y_1),\ldots,\text{sgn}(y_t))~G(|y_1|,\ldots,|y_t|)~f_t^Y(y_1,\ldots,y_t)\\
\nonumber
&=&\int_0^{\infty}dy_1\cdots\int_0^{\infty}dy_t
\sum_{b_1\in\mathbb{Z}_2}\cdots\sum_{b_t\in\mathbb{Z}_2} F(b_1,\ldots,b_t)~G(y_1,\ldots,y_t)~f_t^Y(y_1,\ldots,y_t)\\
\nonumber
&=&2^{-t}\sum_{b_1\in\mathbb{Z}_2}\cdots\sum_{b_t\in\mathbb{Z}_2} F(b_1,\ldots,b_t)~\cdot\\
\nonumber
&\cdot& 2^t\int_0^{\infty}dy_1\cdots\int_0^{\infty}dy_t~G(y_1,\ldots,y_t)~f_t^Y(y_1,\ldots,y_t)\\
&=& 2^{-t}\sum_{b_1\in\mathbb{Z}_2}\cdots\sum_{b_t\in\mathbb{Z}_2} F(b_1,\ldots,b_t)\cdot\Att[G(|Y_1|,\ldots,|Y_t|)]~,
\label{eqB|Y|1}
\end{eqnarray}
where both the second and the last equalities are due to
Eq.\ (\ref{eqFb}) and the symmetry of $f_t^Y$. Plugging here
$G(y_1,\ldots,y_t)=1$ at first, we get Eq.\ (\ref{eqBdist}).
Eq.\ (\ref{eqB|Y|}) is then a consequence of Eqs.\ (\ref{eqBdist}) and
(\ref{eqB|Y|1}).
$\qquad\Box$

\section{Tail behavior of $f_1^X$} 
\label{sec:tail}
In the Main Text we considered the issue about the tail of $f_1^X$ very briefly. 
Here we provide a more extended discussion on this point studying the
expectation $\Att\bigl[|X_1|^q\bigr]$ for $q>0$. 
In addition, we need to consider
similar expectations also below in this Supplementary Material.

The independence between $a_{I_1}$ and $Y_1$ allows us to write
$\Att\bigl[|X_1|^q\bigr]=\Att\bigl[a^q_{I_1}\bigr]\Att\bigl[|Y_1|^q\bigr]$.
Since Eq. (19) of the Main Text for the sequence $\{a_i\}_{i=1}^{\infty}$
makes $\Att\bigl[a^q_{I_1}\bigr]$ always finite, we realize that
$\Att\bigl[|X_1|^q\bigr]$ is finite if and only if
$\Att\bigl[|Y_1|^q\bigr]$ is finite. On the other hand,
$\Att\bigl[|Y_1|^q\bigr]$ is finite if and only if
$\int_{0}^{\infty}\sigma^q\rho(\sigma)d\sigma$ is so. Indeed, from the
definitions of $f_1^Y$ and $\varphi_1$ [Eqs.\ (\PDFYpaper) and
(\phipaper) of the Main Text, respectively] we get
\begin{eqnarray}
\nonumber
\Att\bigl[|Y_1|^q\bigr]&=&\int_{\Real}|x|^q\mathcal N_1(x)\cdot\int_{0}^{\infty}\sigma^q\rho(\sigma)d\sigma\\
&=&\frac{2^{\frac{q}{2}}}{\sqrt{\pi}}\Gamma\biggl(\frac{q+1}{2}\biggr)\int_{0}^{\infty}\sigma^q\rho(\sigma)d\sigma~,
\label{f1q}
\end{eqnarray}
where $\Gamma$ is the Euler's Gamma function.  Thus we discover that
$\Att\bigl[|X_1|^q\bigr]$ and $\Att\bigl[|Y_1|^q\bigr]$, and the tails
of $f_1^X$ as a consequence, are only ruled by the last factor on the r.h.s.
of Eq.\ (\ref{f1q}), i.e. by $\rho$. In particular, if the function
$\rho$ decays according to a power law as $\sigma^{-\alpha-1}$ for
large $\sigma$, then $f_1^X$ inherits the same feature and displays
fat tails with the same tail index $\alpha$.

In the Main Text we also noticed that an effective fat tail scenario
can be obtained also by considering suitable small values of the
restart probability $\nu$ if $D<1/2$. In order to shed light on this
issue, we need to consider the limit of $f_1^X$ when $\nu$ goes to
zero. Nevertheless, such a limit is meaningless if we do not rescale
the function $\rho$ properly with $\nu$ since, if $\rho$ is kept fixed
in the limit procedure, then $f_1^X$ concentrates around zero. The reason
is that, when restarts get very rare, the random time $\I$ tends to
never go back to $1$, 
reaching very large values in equilibrium conditions. As a
consequence, when $D<1/2$ the rescaling factor $a_{I_1}$ tends to
vanish and thus the mixture giving $f_1^X$ becomes 
dominated by component distributions having a vanishing variance.
If we want then to keep the variance of the random variable $X_1$
independent of $\nu$, we must fix a function $\rho'$ which
does not depend on $\nu$ and define
$\rho$ according to 
\begin{equation}
\rho(\sigma)\equiv\sqrt{\Att[a^2_{I_1}]}~\rho'\biggl(\sqrt{\Att[a^2_{I_1}]}~\sigma\biggr)~.
\end{equation}
Indeed, Eq.\ (\ref{f1q}) now provides
\begin{equation}
\Att\bigl[|X_1|^q\bigr]=\Att\bigl[a^q(I_1)\bigr]\Att\bigl[|Y_1|^q\bigr]=
\frac{2^{\frac{q}{2}}}{\sqrt{\pi}}\Gamma\biggl(\frac{q+1}{2}\biggr)\frac{\Att\bigl[a^q(I_1)\bigr]}{\Att\bigl[a^2(I_1)\bigr]^{\frac{q}{2}}}
\int_{0}^{\infty}\sigma^q\rho'(\sigma)d\sigma~,
\label{Xnusmall}
\end{equation}
which in particular entails
\begin{equation}
\Att\bigl[X_1^2\bigr]=\int_{0}^{\infty}\sigma^2\rho'(\sigma)\;d\sigma~.
\end{equation}
In this framework the fluctuations of $X_1$ do not shrink in the limit $\nu\to0$ 
and we obtain
\begin{equation}
\lim_{\nu\to 0^+}\Att\bigl[|X_1|^q\bigr]=\frac{2^{\frac{q}{2}}}{\sqrt{\pi}}
\frac{\Gamma\bigl(\frac{q+1}{2}\bigr)\Gamma\bigl(\frac{2-(1-2D)q}{2}\bigr)}{\Gamma^{\frac{q}{2}}(2D)}\int_{0}^{\infty}\sigma^q\rho'(\sigma)\;d\sigma
\end{equation}
if $(1/2-D)q<1$ and $\lim_{\nu\to 0^+}\Att\bigl[|X_1|^q\bigr]=\infty$
otherwise. 
We thus get the proof that fat tails with index
$2/(1-2D)$ appear in this limit situation.  It is clear that a
function $\rho'$ which endows the model with tails characterized by a
tail index $\alpha<2/(1-2D)$ hides this effect. 

We conclude sketching
the computation of this limit.  To this purpose, 
it is convenient to introduce the
notation of asymptotic equivalence: given two generic
functions $F$ and $G$ of $\nu$, we shall write $F\sim G$ to say that
$\lim_{\nu\to 0^+}F(\nu)/G(\nu)=1$. Then, if
$\{\psi_i\}_{i=1}^{\infty}$ is a sequence for which there exists
$\gamma>0$ and $l>0$ such that $\lim_{i\to\infty}i^{\gamma}\psi_i=l$,
we have that as $\nu$ goes to zero
\begin{equation}
\Att[\psi_{I_1}]=\sum_{i=1}^{\infty}\psi_i\nu(1-\nu)^{i-1}\sim
\begin{cases}
l\Gamma(1-\gamma)\nu^\gamma & \mbox{ if }0<\gamma<1;\\
l\nu|\ln\nu| & \mbox{ if }\gamma=1;\\
\nu\sum_{i=1}^{\infty}\psi_i & \mbox{ if }\gamma>1.
\end{cases}
\label{limnu}
\end{equation}
The last series is convergent. The instance
$\gamma\le 1$ is a consequence of the Karamata's theorem \cite{Feller},
whereas the case $\gamma>1$ is due to the Abel's theorem
\cite{Feller}. The limit value of Eq.\ (\ref{Xnusmall}) follows then
by noticing that $a_i=\sqrt{i^{2D}-(i-1)^{2D}}$ implies
$\lim_{i\to\infty}i^{1/2-D}a_i=\sqrt{2D}$.
$\qquad\Box$

\section{Scaling features}

\label{sec:scaling}
We reconsider here the scaling features of our model at short time
scales, providing a deeper insight into the properties 
outlined in Section IV B of the Main Text. 

To begin with, we derive the distribution of the aggregated
return $X_1+\cdots+X_t$ when $t\le M+1$. To this aim, let us observe
that from Eqs.\ (\ref{ft}) and (\ref{phiY}), thanks to the stability
of Gaussian distributions with respect to linear combinations of 
independent Gaussian variables, we
attain
\begin{equation}
\Att[F(X_1+\cdots+X_t)]=\int_{\Real}dx~F(x)~\Att\biggl[\int_{0}^{\infty}d\sigma~\rho(\sigma)~
\mathcal{N}_{\sqrt{a^2_{I_1}+\cdots+a^2_{I_t}}\sigma}(x)\biggr]
\label{fscaling}
\end{equation}
for any test function $F$ on $\Real$. This identity clearly shows
that, if $t\le M+1$, the PDF of $X_1+\cdots+X_t$ is given as a
function of $x$ by the expression
\begin{equation}
\Att\biggl[\int_{0}^{\infty}d\sigma~\rho(\sigma)~\mathcal{N}_{\sqrt{a^2_{I_1}+\cdots+a^2_{I_t}}\sigma}(x)\biggr]~.
\end{equation}
We notice that this PDF cannot be obtained by simply rescaling $f_1^X$,
except if $\nu=1$ or $a_i=1$ for any $i$ in which case the normal scaling
behavior with exponent $1/2$ is recovered. Thus, in general
the model accounts for a richer scenario 
than a perfect time-scale-invariance framework, as we know.

Choosing in Eq.\ (\ref{fscaling}) $F(x)=|x|^q$, with $q\ge 0$ such that
$\int_{0}^\infty \sigma^q\rho(\sigma)d\sigma<\infty$, we get
\begin{equation}
\Att[|X_1+\cdots+X_t|^q]=\frac{2^{\frac{q}{2}}}{\sqrt{\pi}}\Gamma\biggl(\frac{q+1}{2}\biggr)\int_{0}^\infty \sigma^q\rho(\sigma)d\sigma\cdot
\Att\bigl[\bigl(a^2_{I_1}+\cdots+a^2_{I_t}\bigr)^{\frac{q}{2}}\bigr]~.
\label{Xscaling}
\end{equation}
This result allows us to prove Eq.\ (\mpaper) of the Main Text:
\begin{equation}
m^X_q(t)=\frac{\Att[|X_1+\cdots+X_t|^q]}{\Att[|X_1|^q]}=
\frac{\Att\bigl[\bigl(a^2_{I_1}+\cdots+a^2_{I_t}\bigr)^{\frac{q}{2}}\bigr]}
     {\Att\bigl[a^q_{I_1}\bigr]}~.
\label{phidef}
\end{equation}
Notice in passing
that the r.h.s. of Eq.\ (\ref{phidef}) is well defined for
any real $q$, even if $\Att[|X_1|^q]$ diverges.

\begin{figure}
\centering
\psfrag{q}[ct][ct][2.]{$q$}
\psfrag{epsq}[ct][ct][2.]{$\epsilon_q$}
\includegraphics[angle=-90,scale=0.5]{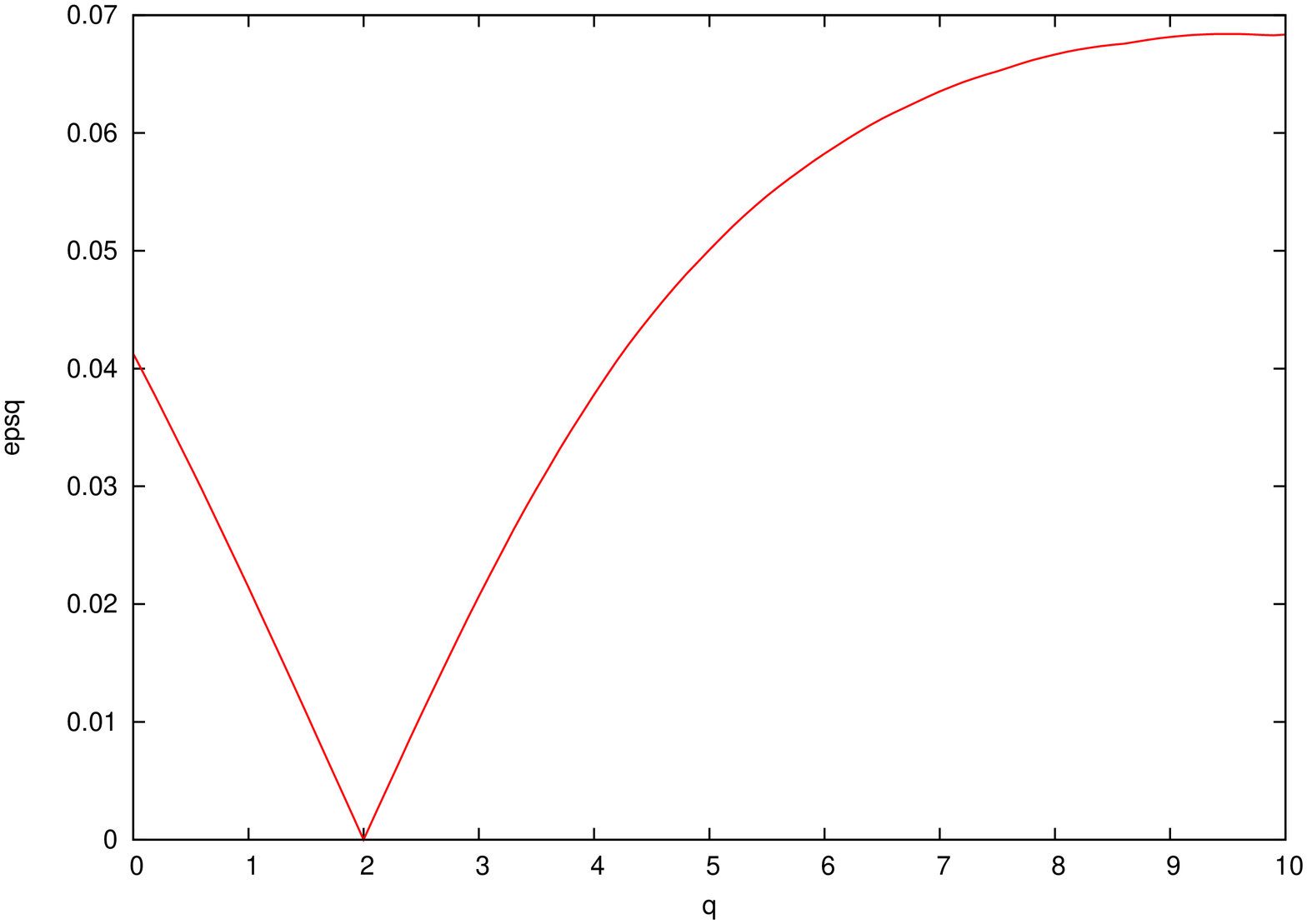}
\caption{Relative deviation of $m^X_q$ from $t^{qH_q}$ vs. $q$. \label{fig:epsq}}
\end{figure}

\begin{figure}
\centering
\psfrag{phiq}[ct][ct][2.]{$m^X_q$}
\psfrag{t}[ct][ct][2.]{$t$}
\psfrag{q=0.5}[ct][ct][1.5]{$q=0.5$}
\psfrag{q=1}[ct][ct][1.5]{$q=1$}
\psfrag{q=3}[ct][ct][1.5]{$q=3$}
\psfrag{q=4}[ct][ct][1.5]{$q=4$}
\includegraphics[angle=-90,scale=0.6]{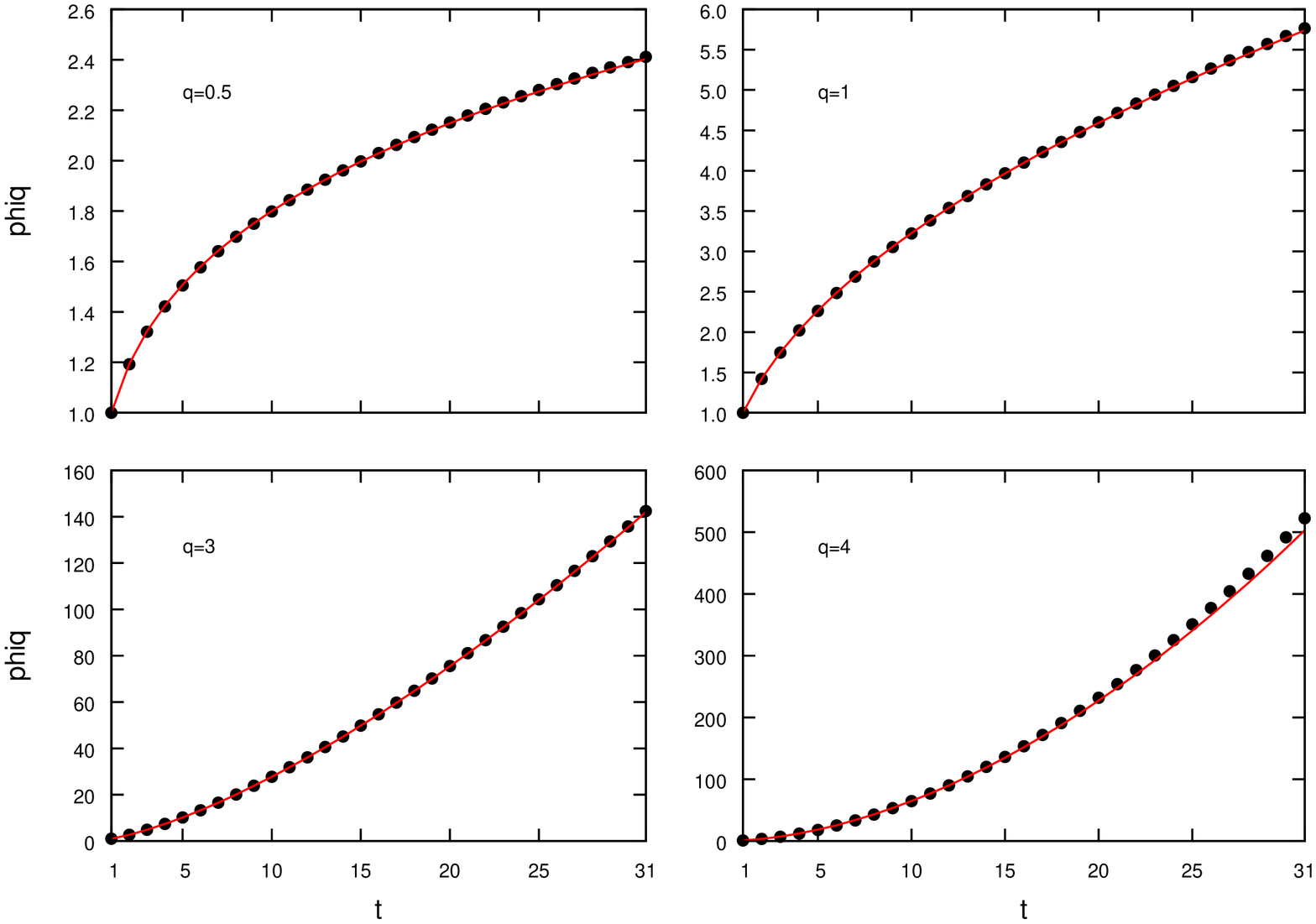}
\caption{Comparison between $m^X_q$ (dots) and $t^{qH_q}$ (dashed lines) for $t\le 31$,
$\nu=0.01$, $D=0.25$ and some values of $q$. \label{fig:somePhi}}
\end{figure}

As we reported in the Main Text, $m^X_q$ is well approximated for not
too small values of $D$ by the power $t^{qH_q}$ with a generalized
Hurst-like exponent $H_q$ independent of $t$, thus allowing
the model to exhibit pretty well-defined scaling properties at 
relatively short
time scales.  
To corroborate this assertion, here we report a study of $m^X_q$ based on
numerical simulations for $M=30$, $0.1\le D\le 1/2$, and $t\le M+1$.  We
evaluate the exponent $H_q$ with the least square method as follow:
\begin{equation}
qH_q\equiv\underset{H\in\Real}{\operatorname{argmin}}\left\{
\sqrt{\frac{1}{M}\sum_{t=2}^{M+1}\biggl[\frac{\ln m^X_q(t)}{\ln t}-qH\biggr]^2}\right\}=
\frac{1}{M}\sum_{t=2}^{M+1}\frac{\ln m^X_q(t)}{\ln t}~.
\label{xiq}
\end{equation}
Then, we measure the distance of $m^X_q$ from $t^{qH_q}$ with the
relative mean fluctuation
\begin{equation}
\epsilon_q\equiv\max\left\{
\frac{1}{qH_q}\sqrt{\frac{1}{M}\sum_{t=2}^{M+1}\biggl[\frac{\ln m^X_q(t)}{\ln t}-qH_q\biggr]^2}
~:~\nu\in[0,1]~\text{and}~D\in[0.1,1/2]\right\}~.
\end{equation}
Even if not explicitly indicated, 
it is clear that $m^X_q$ and $H_q$ depend on $\nu$ and $D$. 
Fig.\ \ref{fig:epsq}
shows $\epsilon_q$ vs. $q$, for $q$ in between 0 and 10.   
The fact that $\epsilon_2=0$ is not surprising since
$m^X_2(t)=t$ as one can immediately verify recalling the stationarity
of $\I$.  For $q\neq2$, we find values of $\epsilon_q$ of few
points per cent. This confirms that $m^X_q$ is close to $t^{qH_q}$ and
motivates the analysis of the exponent $H_q$.  
In Fig.\ \ref{fig:somePhi} 
we report an explicit comparison between $m^X_q$ and $t^{qH_q}$
for $q=0.5,~1,~3,~4$, $\nu=0.01$, and $D=0.25$. The corresponding 
$H_q$ vs. $q$ plot is shown in Fig.\ 1 of the Main Text.

Before we proceed to further investigate $H_q$, a remark is in order about 
small values of $D$.  When $D$ goes to 0, $a_i$ vanishes if $i>1$, thus
approaching $\delta_{i1}$. On the other hand,
$\delta_{I_11},\ldots,\delta_{I_t1}$ are independent identically
distributed Bernoulli variables, since in our model there is no
correlations between restarts. Then, one can easily verify that in
such a limit $m^X_q(t)$ takes the simple form
\begin{equation}
\sum_{n=1}^t n^{\frac{q}{2}} {t \choose n}\nu^{n-1}(1-\nu)^{t-n}~.
\end{equation}
This function is poorly approximated by a power of the time when $q$
is small and $\nu$ assumes intermediate values: for instance, with
$q=0$ and $\nu=1/2$, it reduces to $2(1-2^{-t})$. This is the reason
that leads us to exclude small value of $D$ and to focus on $D\ge 0.1$
in the scaling analysis. It is also worth noticing that
$\nu$ is typically much smaller than $1/2$ in order 
to reproduce empirical financial data.

\begin{figure}
\centering
\psfrag{D}[ct][ct][1.]{$D$}
\psfrag{nu}[ct][ct][1.]{$\nu$}
\includegraphics[angle=-90,scale=1]{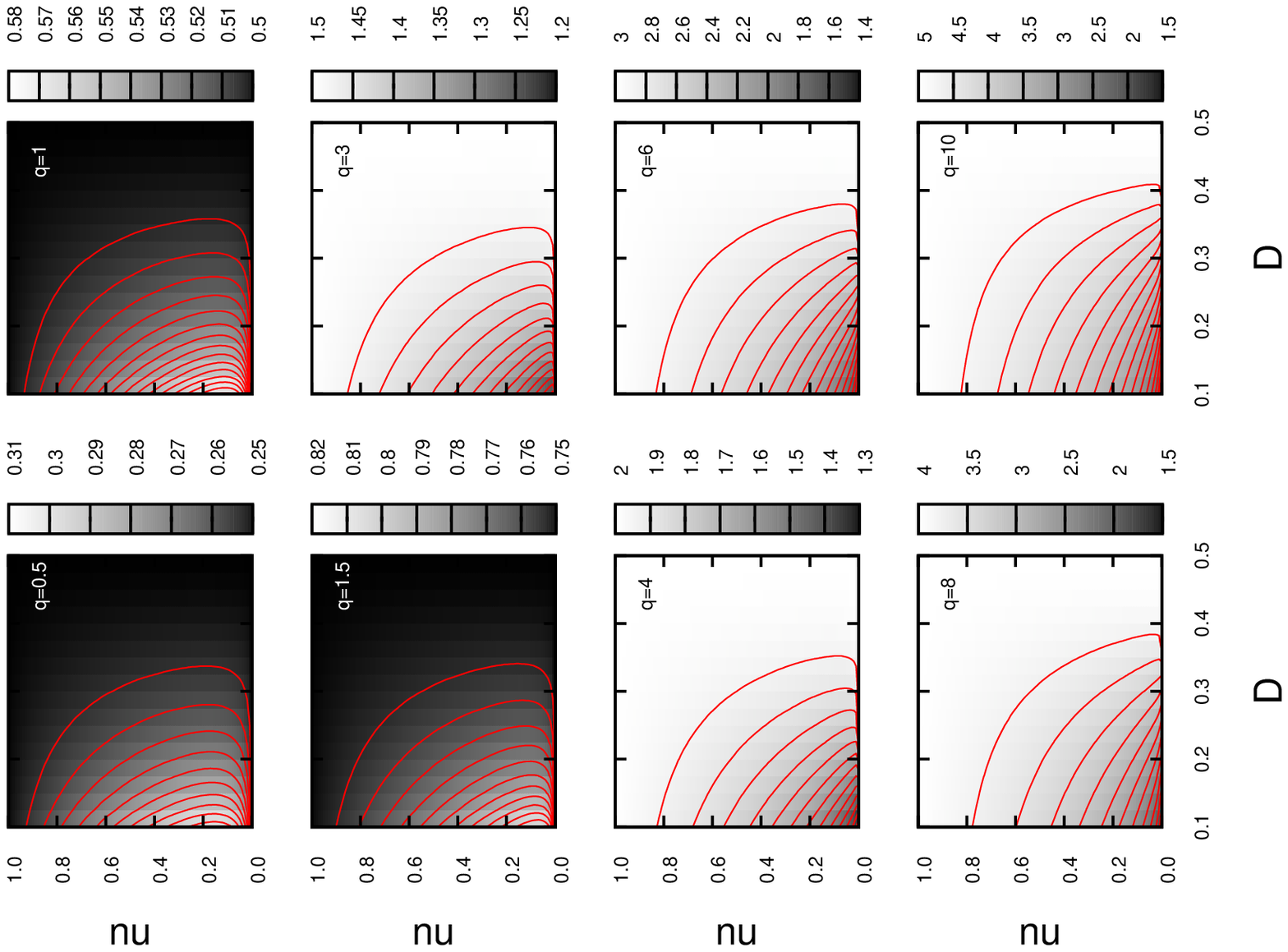}
\caption{Scaling exponent $qH_q$ as a function of $\nu$ and $D$ for several $q$. 
Contour lines are also shown. \label{fig:xiq}}
\end{figure}

For the scaling exponent $H_q$ defined by Eq.\ (\ref{xiq}), we have
$H_q\geq1/2$ if $q\le 2$ and $H_q\leq1/2$
if $q>2$. Indeed, the function
\begin{equation}
F(x_1,\ldots,x_t)\equiv\biggl(x_1^{\frac{1}{\alpha}}+\cdots+x_t^{\frac{1}{\alpha}}\biggr)^\alpha
\end{equation}
is convex when $\alpha\le 1$ and concave if $\alpha>1$. Thus, setting
$\alpha=q/2$, the Jensen's inequality and the stationarity of the
process $\I$ tell us that if $q\le 2$ then
\begin{equation}
m^X_q(t)=\frac{\Att\bigl[F\bigl(a^q_{I_1},\ldots,a^q_{I_t}\bigr)\bigr]}{\Att\bigl[a^q_{I_1}\bigr]}\ge
\frac{F\bigl(\Att\bigl[a^q_{I_1}\bigr],\ldots,\Att\bigl[a^q_{I_t}\bigr]\bigr)}{\Att\bigl[a^q_{I_1}\bigr]}=t^{\frac{q}{2}},
\end{equation}
while for $q>2$
\begin{equation}
m^X_q(t)=\frac{\Att\bigl[F\bigl(a^q_{I_1},\ldots,a^q_{I_t}\bigr)\bigr]}{\Att\bigl[a^q_{I_1}\bigr]}\le
\frac{F\bigl(\Att\bigl[a^q_{I_1}\bigr],\ldots,\Att\bigl[a^q_{I_t}\bigr]\bigr)}{\Att\bigl[a^q_{I_1}\bigr]}=t^{\frac{q}{2}}~.
\end{equation}
The bounds on $H_q$ then follow by its definition, Eq.\ (\ref{xiq}).
Fig.\ \ref{fig:xiq} shows the level curves of $qH_q$ vs. $\nu$ and $D$ 
in the range $[0,1]$ and $[0.1,1/2]$, respectively, and for
different values of $q$. The exponent $H_q$ displays large variations
for small values of $\nu$ and $D$ when $q>2$ whereas it is close to
$1/2$ in the other cases. Moreover, given $\nu$ we notice that $H_q$
is a decreasing function of $D$ if $q<2$ and an increasing function if
$q>2$. Disentangling the contribution of $\nu$ and $D$ to the scaling
exponent is not easy. However, from the contour lines in 
Fig.\ \ref{fig:xiq} it is possible to appreciate
how the variations of one of the two
parameters can be compensated by modifications of the other.

We conclude the present analysis by studying analytically the limit of
$m^X_q$ when $D\le 1/2$ and $\nu$ approaches zero. We shall make use of
the symbol $\sim$ of asymptotic equivalence introduced in
the previous Section and the results in Eq.\ (\ref{limnu}). The
computation starts by isolating the trajectories corresponding to at most
one restart from the others in the numerator of $m^X_q$. Then,
remembering $a^2_i=i^{2D}-(i-1)^{2D}$ and assuming $t\ge 2$,
we get the first equivalence
\begin{eqnarray}
\nonumber
\Att\bigl[\bigl(a^2_{I_1}+\cdots+a^2_{I_t}\bigr)^{\frac{q}{2}}\bigr]&\sim&\sum_{i=1}^{\infty}\bigl[(i+t-1)^{2D}-(i-1)^{2D}\bigr]^{\frac{q}{2}}\nu(1-\nu)^{i+t-2}\\
&+&\sum_{\tau=2}^t\sum_{i=1}^\infty\bigl[(i+\tau-1)^{2D}-(i-1)^{2D}+(t-\tau+1)^{2D}\bigr]^{\frac{q}{2}}\nu^2(1-\nu)^{i+t-3}~.
\label{limphi}
\end{eqnarray}
Noticing now that
\begin{equation}
\lim_{i\to\infty}\bigl[(i+\tau-1)^{2D}-(i-1)^{2D}+(t-\tau+1)^{2D}\bigr]^{\frac{q}{2}}=(t-\tau+1)^{Dq}
\end{equation}
and that $(t-\tau+1)^{Dq}$ does not vanish for $\tau=2,\ldots,t$, we
see that the second term in Eq. (\ref{limphi}) is
$\sim\nu\sum_{\tau=1}^{t-1}\tau^{Dq}$. Thus, Eq.\ (\ref{limphi}) can
be recast as
\begin{equation}
\Att\bigl[\bigl(a^2_{I_1}+\cdots+a^2_{I_t}\bigr)^{\frac{q}{2}}\bigr]\sim\sum_{i=1}^{\infty}\bigl[(i+t)^{2D}-i^{2D}\bigr]^{\frac{q}{2}}\nu(1-\nu)^{i+t-1}
+\nu\sum_{\tau=1}^t\tau^{Dq}~,
\label{scalingsmallnu}
\end{equation}
which is valid also for $t=1$. To obtain this relation we shifted the
index $i$ in the first sum of Eq.\ (\ref{limphi}) and then we moved
the first addend to the second term. Finally, since
$\lim_{i\to\infty}i^{(1/2-D)q}\bigl[(i+t)^{2D}-i^{2D}\bigr]^{\frac{q}{2}}=(2Dt)^{\frac{q}{2}}$,
Eq.\ (\ref{limnu}) allows us to obtain
\begin{equation}
m^X_q(t)\sim
\begin{cases}
t^{\frac{q}{2}}  & \mbox{ if }(1/2-D)q\le 1;\\
\frac{\displaystyle{\sum_{i=1}^{\infty}\bigl[(i+t)^{2D}-i^{2D}\bigr]^{\frac{q}{2}}
+\sum_{\tau=1}^t\tau^{Dq}}}{\displaystyle{\sum_{i=1}^{\infty}\bigl[(i+1)^{2D}-i^{2D}\bigr]^{\frac{q}{2}}+1}} &  \mbox{ if }(1/2-D)q>1.
\label{scalsmallnu}
\end{cases}
\end{equation}
This asymptotic equivalence proves that $H_q=1/2$ for $q\le 2/(1-2D)$
at small values of $\nu$, as anticipated in the Main Text. In order to
intuitively understand the content of such a result, we notice that
when the restart probability $\nu$ goes to zero the random time $\I$
tends to flow without stopping but its starting value becomes
affected by very large fluctuations due to its stationarity. At
small values of $q$, the sequence $\{a^q_i\}_{i=1}^{\infty}$ does not
decay fast enough to keep under control such fluctuations and only its
tail plays a role, providing a normal scaling exponent.

\section{The ``null'' and the ``complete'' model}
\label{sec:simplerho}
In the Main Text we considered two particular choices for the function
$\rho$: one corresponded to fix $\sigma$ to a particular value and,
being the simplest possible choice, we referred to the
``null model'' in that case; and the other, given by Eq.\ (\rhopaper) of the Main Text, was
obtained distributing $\sigma^2$ according to an inverse-gamma
distribution.  The model associated to the latter choice for $\rho$ was named the
complete model.  Here we give some details about the null
model and prove that the
endogenous component of the complete model is an ARCH process, as
stated in the Main Text.

\subsection{The null model}

When $\sigma$ is fixed to a particular value $\sigma_0$, the PDF
$\varphi_t$ of Eq.\ (\ref{phiY}) factors and, consequently, from
Eq.\ (\ref{gesplic}) we have that the joint PDF $f^Y_t$ also factors as
\begin{equation}
f^Y_t(y_1,\ldots,y_t)=\prod_{n=1}^t\mathcal{N}_{\sigma_0}(y_n)~.
\end{equation}
Within this setting, the endogenous component $\Y$ then reduces to a
sequence of independent normal variables, which is the simplest
possible endogenous process that our model can produce, and
the observed compound process becomes a random time change of the
Brownian motion. 
Thus, we are recovering 
a discrete-time model
with
random time and constant average volatility.  
Indeed, without demanding mathematical
rigor, if $(W_t)_{t\ge 0}$ is a standard Brownian motion independent
of $\I$, then $X_1+\cdots+X_t$ is distributed as
$W_{\sigma_0^2[a^2_{I_1}+\cdots+a^2_{I_t}]}$ for any $t$ when $\Y$ is a
sequence of i.i.d. normal variables with mean zero and variance
$\sigma_0^2$.

As we discussed in the Main Text and in Section \ref{sec:tail}, if
$D<1/2$ we can have effective fat tails with tail index $2/(1-2D)$ in
the distribution of $X_1$ by considering a small enough value of the
restart probability $\nu$. This is the only possibility to obtain such
tails within the present instance of the model.  The rescaling of
$\rho$ we considered in Section \ref{sec:tail} simply consists in
taking $\sigma_0=\overline\sigma/\sqrt{\Att[a^2_{I_1}]}$ here, with
$\overline\sigma$ a parameter independent of $\nu$.

\subsection{The complete model}

The peculiar $\rho$ given in the Main Text by Eq.\ (\rhopaper) allows us
to explicitly integrate over $\sigma$ in the expression of $\varphi_t$,
which reduces to a multivariate Student distribution:
\begin{equation}
\varphi_t(y_1,y_2,\ldots,y_t)=\frac{\Gamma(\frac{\alpha+t}{2})}{(\sqrt{\pi}\beta)^t\Gamma(\frac{\alpha}{2})}
\biggl[1+\frac{y_1^2+y_2^2+\cdots +y_t^2}{\beta^2}\biggr]^{-\frac{\alpha+t}{2}}~.
\label{phisimple}
\end{equation} 
Within this setting, reformulating the endogenous component $\Y$ in
terms of stochastic variables, rather than only stating its PDF's, is
interesting and useful. We write such process as
\begin{equation}
Y_t=\begin{cases}
\beta\cdot Z_1 & \mbox{if } t=1;\\
\sqrt{\beta^2+Y_{\max\{1,t-M\}}^2+\cdots Y_{t-1}^2}\cdot Z_t & \mbox{if } t>1,
\label{ARCHY}
\end{cases}
\end{equation}
with a residual sequence $\{Z_t\}_{t=1}^\infty$ obviously defined as
\begin{equation}
Z_t=\begin{cases}
Y_1/\beta & \mbox{if } t=1;\\
Y_t/\sqrt{\beta^2+Y_{\max\{1,t-M\}}^2+\cdots Y_{t-1}^2} & \mbox{if } t>1.
\end{cases}
\label{residual}
\end{equation}
Here we show that $\{Z_t\}_{t=1}^\infty$ is a sequence of Student's
t-distributed independent variables when $\varphi_t$ is given by
Eq.\ (\ref{phisimple}). According to the definition of the ARCH process \cite{tsay_1},
this fact makes $\Y$ a pure ARCH process with Student's t-distributed
return residuals, as anticipated in the Main Text. To be more precise, we
can prove that
\begin{equation}
f_t^{Z}(z_1,\ldots,z_t)=\prod_{n=1}^t\frac{\Gamma(\frac{\alpha_n+1}{2})}{\sqrt{\pi}\Gamma(\frac{\alpha_n}{2})}
(1+z_n^2)^{-\frac{\alpha_n+1}{2}}~,
\label{Zdist}
\end{equation}
with $\alpha_n\equiv\alpha+\min\{n-1,M\}$. Notice that $Z_t$'s are 
identically distributed for $t\ge M+1$, but the stationarity of $\Y$
and the boundary effects at $t=1$ prevent them to be identically distributed
for any $t$. It is also worth mentioning that simulating the process
$\Y$ becomes rather simple thanks to the algorithm reported in Ref.\
\cite{bailey}, which adapts the Box-Muller transform for normally
distributed variables to Student's t-distributed variables.

In order to prove Eq.\ (\ref{Zdist}) we study the expectation value
$\Att[F(Z_1,Z_2,\ldots,Z_t)]$, being $F$ a test function on $\Real^t$.
By definition, Eq.\ (\ref{residual}), we have
\begin{eqnarray}
\nonumber \Att[F(Z_1,Z_2,\ldots,Z_t)]
&=&\Att\biggl[F\biggl(Y_1/\beta,Y_2/\sqrt{\beta^2+Y_1^2},\ldots,Y_t/\sqrt{\beta^2+Y_{\max\{1,t-M\}}^2+\cdots
Y_{t-1}^2}\biggr)\biggr]\\ \nonumber
&=&\int_{\Real}dy_1\int_{\Real}dy_2\cdots\int_{\Real}dy_t~
F\left(y_1/\beta,y_2/\sqrt{\beta^2+y_1^2},\ldots,y_t/\sqrt{\beta^2+y_{\max\{1,t-M\}}^2+\cdots
y_{t-1}^2}\right)\cdot\\ 
&\cdot& f_t^{Y}(y_1,y_2,\ldots,y_t)~.
\end{eqnarray}
We then perform a change of variables from the old $y_n$'s into the
new
\begin{equation}
z_n=\begin{cases}
y_1/\beta & \mbox{if } n=1;\\
y_n/\sqrt{\beta^2+y_{\max\{1,n-M\}}^2+\cdots y_{n-1}^2} & \mbox{if } 1<n\le t.
\end{cases}
\label{trasf}
\end{equation}
This relation can be inverted to express the $y_n$'s as a function of
the $z_n$'s.  Since clearly $y_n$ only depends on
$z_1,z_2,\ldots,z_n$, the Jacobian matrix of the transformation is
triangular and thus its determinant is easily found as
\begin{equation}
\beta\prod_{n=2}^t\sqrt{\beta^2+y_{\max\{1,n-M\}}^2+\cdots y_{n-1}^2}~.
\end{equation}
Here and below the $y_n$'s must be thought as functions of the 
$z_n$'s.  Such a change of variables leads us to the identity
\begin{eqnarray}
\Att[F(Z_1,Z_2,\ldots,Z_t)]
\nonumber
&=&\int_{\Real}dz_1\int_{\Real}dz_2\cdots\int_{\Real}dz_t~F(z_1,z_2,\ldots,z_t)\cdot\\
&\cdot&\beta\prod_{n=2}^t\sqrt{\beta^2+y_{\max\{1,n-M\}}^2+\cdots y_{n-1}^2}~
f_t^Y(y_1,y_2,\ldots,y_t)~.
\end{eqnarray}
Now, although lengthy, using Eqs.\ (\ref{gesplic}) and
(\ref{phisimple}) at first and Eq.\ (\ref{trasf}) at last, it is
straightforward to see that
\begin{eqnarray}
\nonumber
&&\beta\prod_{n=2}^t\sqrt{\beta^2+y_{\max\{1,n-M\}}^2+\cdots y_{n-1}^2}~f_t^Y(y_1,y_2,\ldots,y_t)\\
\nonumber
&=&\frac{\Gamma(\frac{\alpha+1}{2})}{\sqrt{\pi}\Gamma(\frac{\alpha}{2})}\biggl[1+\frac{y_1^2}{\beta^2}\biggr]^{-\frac{\alpha+1}{2}}
\prod_{n=2}^t\frac{\Gamma(\frac{\alpha_n+1}{2})}{\sqrt{\pi}\Gamma(\frac{\alpha_n}{2})}
\biggl[1+\frac{y_n^2}{\beta^2+y_{\max\{1,n-M\}}^2+\ldots+y_{n-1}^2}\biggr]^{-\frac{\alpha_n+1}{2}}\\
&=&\prod_{n=1}^t\frac{\Gamma(\frac{\alpha_n+1}{2})}{\sqrt{\pi}\Gamma(\frac{\alpha_n}{2})}
(1+z_n^2)^{-\frac{\alpha_n+1}{2}}~,
\end{eqnarray}
where $\alpha_n\equiv\alpha+\min\{n-1,M\}$.
Hence
\begin{equation}
\Att[F(Z_1,Z_2,\ldots,Z_t)]
=\int_{\Real}dz_1\int_{\Real}dz_2\cdots\int_{\Real}dz_t~F(z_1,z_2,\ldots,z_t)~
\prod_{n=1}^t\frac{\Gamma(\frac{\alpha_n+1}{2})}{\sqrt{\pi}\Gamma(\frac{\alpha_n}{2})}
(1+z_n^2)^{-\frac{\alpha_n+1}{2}}
\end{equation}
and this eventually confirms that the process $\{Z_t\}_{t=1}^{\infty}$ is
distributed according to Eq.\ (\ref{Zdist}).
$\qquad\Box$

\section{Autocorrelation structure}
\label{sec:correlation}
Here  we discuss in some detail the autocorrelation $r_q^X$
of the process $\{|X_t|^q\}_{t=1}^{\infty}$ introduced in the
Main Text. To begin with, we notice that the independence between $\Y$ and
$\I$ allows us to write
\begin{equation}
r_q^X(t)\equiv\frac{\Att\bigl[|X_1|^q|X_t|^q\bigr]-\Att\bigl[|X_1|^q\bigr]^2}{\Att\bigl[|X_1|^{2q}\bigr]-\Att\bigl[|X_1|^q\bigr]^2}
=\frac{\Att\bigl[a^q_{I_1} a^q_{I_t}\bigr]\Att\bigl[|Y_1|^q|Y_t|^q\bigr]-\Att\bigl[a^q_{I_1}\bigr]^2\Att\bigl[|Y_1|^q\bigr]^2}
{\Att\bigl[a^{2q}_{I_1}\bigr]\Att\bigl[|Y_1|^{2q}\bigr]-\Att\bigl[a^q_{I_1}\bigr]^2\Att\bigl[|Y_1|^q\bigr]^2}~.
\label{rX}
\end{equation}
Although
expectations involving $\{a_{I_t}\}_{t=1}^{\infty}$ are finite for any
$q$, we restrict on values of $q$ such that
$\int_{0}^{\infty}\sigma^{2q}\rho(\sigma)d\sigma<\infty$ in order to
ensure that also those involving $\Y$ are finite. 
The analysis of $r_q^X$ we propose is based on the preliminary
study of the autocorrelations $r_q^{a_I}$ and $r_q^Y$ of the processes
$\{a^q_{I_t}\}_{t=1}^{\infty}$ and $\{|Y_t|^q\}_{t=1}^{\infty}$,
respectively, which is the subject of the next two paragraphs. 
Eventually, we bring together the results to go back over $r_q^X$.  
We deal first with $r_q^{a_I}$.

\subsection{The autocorrelation $r_q^{a_I}$}
The autocorrelation $r_q^{a_{I}}$ can be conveniently manipulated once
one knows the probability of $I_t=j$, given that $I_1=i$. When $j<t$
the event $I_t=j$ occurs only as a consequence of a restart at the
time $t-j+1$ and no restarts during the following $j-1$ steps,
regardless of the value of $I_1$. Thus,
$\Prob[I_t=j|I_1=i]=\nu(1-\nu)^{j-1}$ if $j<t$. On the contrary, when
$j\ge t$, the event $I_t=j$ is only possible if no restart occurs
during the whole temporal interval up to time $t$, since a restart in
between 1 and $t$ would provide a value of $I_t$ smaller than $t$. In
such circumstances $I_t=I_1+t-1$ and then
$\Prob[I_t=j|I_1=i]=(1-\nu)^{t-1}\delta_{j~i+t-1}$ if $j\ge t$. Thus,
\begin{eqnarray}
\nonumber
\Prob[I_t=j|I_1=i]&=&
\begin{cases}
\nu (1-\nu)^{j-1} & \mbox{ if }j<t;\\
(1-\nu)^{t-1}\delta_{j~i+t-1} & \mbox{ if }j\ge t
\end{cases}\\
&=&\Prob[I_1=j]+
\begin{cases}
0 & \mbox{ if }j<t;\\
(1-\nu)^{t-1}\delta_{j~i+t-1}-\nu (1-\nu)^{j-1} & \mbox{ if }j\ge t.
\end{cases}
\label{conditionalI}
\end{eqnarray}

Coming back to $r_q^{a_{I}}$ and fixing $t\ge 2$, we notice that
\begin{eqnarray}
\nonumber
\Att\bigl[a^q_{I_1} a^q_{I_t}\bigr]-\Att\bigl[a_{I_1}^q\bigr]^2&=&
\sum_{i=1}^\infty\sum_{j=1}^\infty a^q_ia^q_j\biggl(\Prob[I_t=j,I_1=i]-\Prob[I_1=j]\cdot\Prob[I_1=i]\biggr)\\
\nonumber
&=&\sum_{i=1}^\infty\sum_{j=1}^\infty a^q_ia^q_j\biggl(\Prob[I_t=j|I_1=i]-\Prob[I_1=j]\biggr)\Prob[I_1=i]\\
&=&\sum_{i=1}^\infty\sum_{j=t}^\infty a^q_ia^q_j\biggl((1-\nu)^{t-1}\delta_{j~i+t-1}-\nu(1-\nu)^{j-1}\biggr)\Prob[I_1=i]~,
\end{eqnarray}
where the result stated in Eq.\ (\ref{conditionalI}) has been used to
get the last equality. Then,
\begin{eqnarray}
\nonumber
\Att\bigl[a^q_{I_1} a^q_{I_t}\bigr]-\Att\bigl[a_{I_1}^q\bigr]^2
&=&(1-\nu)^{t-1}\sum_{i=1}^\infty a^q_ia^q_{i+t-1}\Prob[I_1=i]-\Att\bigl[a_{I_1}^q\bigr]\sum_{j=t}^{\infty}a^q_j\nu(1-\nu)^{j-1}\\
\nonumber
&=&(1-\nu)^{t-1}\Att\bigl[a^q_{I_1}a^q_{I_1+t-1}\bigr]-\Att\bigl[a_{I_1}^q\bigr]\sum_{i=1}^{\infty}a^q_{i+t-1}\nu(1-\nu)^{i+t-2}\\
&=&(1-\nu)^{t-1}\biggl(\Att\bigl[a^q_{I_1}a^q_{I_1+t-1}\bigr]-\Att\bigl[a_{I_1}^q\bigr]\Att\bigl[a^q_{I_1+t-1}\bigr]\biggr)~,
\label{corra}
\end{eqnarray}
the second equality being obtained through the substitution $j=i+t-1$
in the second series. In summary, for the autocorrelation $r_q^{a_I}$ we find
the more manageable expression:
\begin{eqnarray}
\nonumber
r_q^{a_I}(t)&=&\frac{\Att\bigl[a^q_{I_1} a^q_{I_t}\bigr]-\Att\bigl[a^q_{I_1}\bigr]^2}{\Att\bigl[a^{2q}_{I_1}\bigr]-\Att\bigl[a^q_{I_1}\bigr]^2}\\
&=&(1-\nu)^{t-1}\frac{\Att\bigl[a^q_{I_1} a^q_{I_1+t-1}\bigr]-\Att\bigl[a^q_{I_1}\bigr]\Att\bigl[a^q_{I_1+t-1}\bigr]}
{\Att\bigl[a^{2q}_{I_1}\bigr]-\Att\bigl[a^q_{I_1}\bigr]^2}~,
\label{rqa}
\end{eqnarray}
which is valid also for $t=1$ and allows us to investigate the
correlation decay.

\begin{figure}
\centering
\psfrag{t}[ct][ct][2.]{$t$}
\psfrag{ra}[ct][ct][2.]{$r_q^{a(I)}$}
\psfrag{q=1}[ct][ct][1.5]{$q=1$}
\psfrag{q=2}[ct][ct][1.5]{$q=2$}
\psfrag{q=3}[ct][ct][1.5]{$q=3$}
\psfrag{q=4}[ct][ct][1.5]{$q=4$}
\includegraphics[angle=0,scale=0.5]{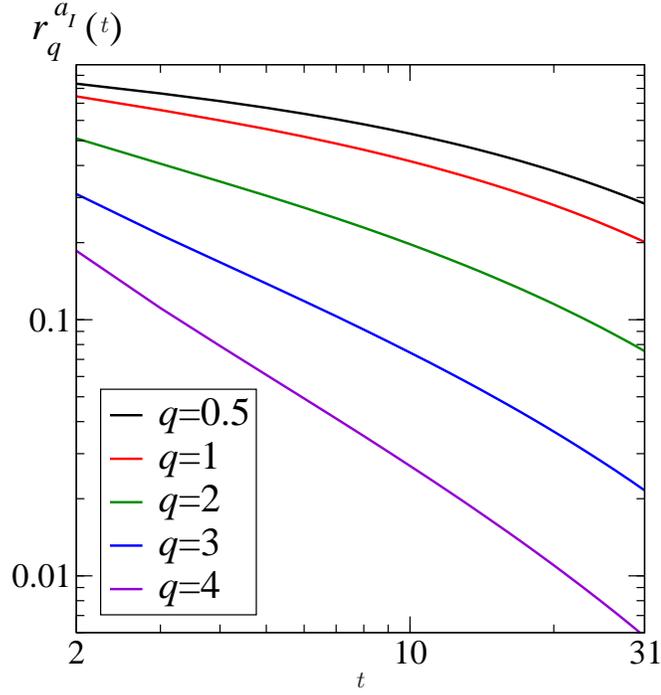}
\caption{Log--log plot of the short time autocorrelation of
  $\{a^q(I_t)\}_{t=1}^{\infty}$ for $\nu=0.01$ and
  $D=0.25$. \label{fig:corra}}
\end{figure}

We point out that $r_q^{a_I}$ decays approximately
according to a power law as $t$ increases at short time scales if
$D<1/2$, whereas in the long time limit an exponential
relaxation with rate $-\ln(1-\nu)$ dominates:
\begin{equation}
\lim_{t\to\infty}\frac{1}{t}\ln r_q^{a_I}(t)=\ln(1-\nu)~.
\end{equation}
The former becomes the main trend at short time
scales and small $\nu$. For instance, Fig.\ \ref{fig:corra} reports
$r_q^{a_I}$ vs. $t$ for $2\le t\le 31$, $\nu=0.01$ and $D=0.25$
and for four different values of $q$. 
Moreover, for $q\le 1/(1-2D)$, the smaller is $\nu$, the more the
correlations get persistent, as we already mentioned. We can
understand this fact by looking at the limit behavior of $r_q^{a_I}$
when $\nu$ approaches zero. Making use of the symbol of asymptotic
equivalence introduced in Section \ref{sec:tail}, we find
\begin{equation}
r_q^{a_I}(t)\sim
\begin{cases}
1 & \mbox{ if } (1-2D)q\le 1;\\
\displaystyle{\frac{\sum_{i=1}^{\infty}a^q_ia^q_{i+t-1}}{\sum_{i=1}^{\infty}a^{2q}_i}} & \mbox{ if } (1-2D)q>1.\\
\end{cases}
\label{ralim}
\end{equation}
This result is an immediate consequence of Eqs.\ (\ref{rqa}) and
(\ref{limnu}). As for the limit scaling behavior, the large
fluctuations of $I_1$ affect $a^q_{I_1}$ when $q$ is small enough. 
In addition, they also propagate to $a^q_{I_1+t-1}$ keeping intact the
correlations.

\subsection{The autocorrelation $r_q^Y$}

The autocorrelation $r_q^Y$ of the process
$\{|Y_t|^q\}_{t=1}^{\infty}$ is much more difficult to investigate
than the previous and in general we must resort to numerical
simulations. Nevertheless it has a trivial structure for $t\le
M+1$. Indeed, due to the exchangeability of $f_t^Y=\varphi_t$ if $t\le
M+1$, we have that
$\Att\bigl[|Y_1|^q|Y_t|^q\bigr]=\Att\bigl[|Y_1|^q|Y_2|^q\bigr]$ when
$2\le t\le M+1$ and thus $r_q^Y$ is independent of the time in such a
temporal interval:
\begin{equation}
r_q^Y(t)=\frac{\Att\bigl[|Y_1|^q|Y_2|^q\bigr]-\Att\bigl[|Y_1|^q\bigr]^2}{\Att\bigl[|Y_1|^{2q}\bigr]-\Att\bigl[|Y_1|^q\bigr]^2}~.
\end{equation}

To the purpose of analyzing $r_q^Y(t)$ also for $t>M+1$, 
we refer here to a favorable instance, corresponding to
the function $\rho$ given by Eq.\ (\rhopaper) of the Main Text with
$\alpha>4$ and $q=2$. As we know, such a $\rho$ makes $\Y$ an ARCH
process for which the condition $\alpha>4$ guarantees the existence of
$\Att\bigl[|Y_1|^4\bigr]$.  We can then take advantage of the fact
that, for ARCH processes, the expectation
$\Att\bigl[F(Y_1)Y_t^2\bigr]$ can be recursively computed for any $t$
and any test function $F$ for which it makes sense.  Indeed, recalling
Eq.\ (\ref{ARCHY}), for $t\ge 2$ we can write
\begin{equation}
Y_t^2=\biggl(\beta^2+\sum_{n=1}^{\min\{t-1,M\}}Y_{t-n}^2\biggr)Z_t^2~.
\end{equation}
Thus, noticing that $Y_n$ is independent of $Z_t$ if $n<t$ and bearing
in mind that $Y_1=\beta Z_1$, we have the simple chain of equalities
\begin{eqnarray}
\nonumber
\Att\bigl[F(Y_1)Y_t^2\bigr]-\Att[F(Y_1)]\cdot\Att\bigl[Y_1^2\bigr]&=&
\Att\biggl[F(Y_1)\biggl(\beta^2+\sum_{n=1}^{\min\{t-1,M\}}Y_{t-n}^2\biggr)Z_t^2\biggr]-\Att[F(Y_1)]\cdot\Att\bigl[Y_1^2\bigr]\\
\nonumber
&=&\Att\biggl[F(Y_1)\biggl(\beta^2+\sum_{n=1}^{\min\{t-1,M\}}Y_{t-n}^2\biggr)\biggr]\cdot\Att\bigl[Z_t^2\bigr]-\Att[F(Y_1)]\cdot\Att\bigl[Y_1^2\bigr]\\
\nonumber
&=&\Att[F(Y_1)]\biggl(\beta^2\Att\bigl[Z_t^2\bigr]-\Att\bigl[Y_1^2\bigr]\biggr)+
\Att\bigl[Z_t^2\bigr]\cdot\sum_{n=1}^{\min\{t-1,M\}}\Att\bigl[F(Y_1)Y_{t-n}^2\bigr]\\
\nonumber
&=&\Att[F(Y_1)]\biggl(\beta^2\Att\bigl[Z_t^2\bigr]-\Att\bigl[Y_1^2\bigr]+\min\{t-1,M\}\Att\bigl[Z_t^2\bigr]\Att\bigl[Y_1^2\bigr]\biggr)+\\
\nonumber
&+&\Att\bigl[Z_t^2\bigr]\cdot\sum_{n=1}^{\min\{t-1,M\}}\biggl(\Att\bigl[F(Y_1)Y_{t-n}^2\bigr]-\Att[F(Y_1)]\cdot\Att\bigl[Y_1^2\bigr]\biggr)\\
\nonumber
&=&\beta^2\Att[F(Y_1)]\biggl(\Att\bigl[Z_t^2\bigr]-\Att\bigl[Z_1^2\bigr]+\min\{t-1,M\}\Att\bigl[Z_t^2\bigr]\Att\bigl[Z_1^2\bigr]\biggr)+\\
&+&\Att\bigl[Z_t^2\bigr]\cdot\sum_{n=1}^{\min\{t-1,M\}}\biggl(\Att\bigl[F(Y_1)Y_{t-n}^2\bigr]-\Att[F(Y_1)]\cdot\Att\bigl[Y_1^2\bigr]\biggr)~.
\label{AAA}
\end{eqnarray}
On the other hand, for the variables $Z_t$'s distributed according to
Eq.\ (\ref{Zdist}) we have
 \begin{equation}
\Att\bigl[|Z_t|^q\bigr]=\frac{\Gamma\bigl(\frac{q+1}{2}\bigr)\Gamma\bigl(\frac{\alpha_t-q}{2}\bigr)}
{\sqrt{\pi}\Gamma\bigl(\frac{\alpha_t}{2}\bigr)}~,
\label{Zq}
\end{equation}
with $\alpha_t=\alpha+\min\{t-1,M\}$. Thus, to verify that
\begin{equation}
\Att\bigl[Z_t^2\bigr]-\Att\bigl[Z_1^2\bigr]+\min\{t-1,M\}\Att\bigl[Z_t^2\bigr]\Att\bigl[Z_1^2\bigr]=0
\label{BBB}
\end{equation}
is not difficult once one sets $q=2$ in Eq.\ (\ref{Zq}). Combining
Eq.\ (\ref{AAA}) with Eq.\ (\ref{BBB}), we eventually obtain the
result
\begin{equation}
\Att\bigl[F(Y_1)Y_t^2\bigr]-\Att[F(Y_1)]\cdot\Att\bigl[Y_1^2\bigr]
=\frac{1}{\alpha_t-2}\sum_{n=1}^{\min\{t-1,M\}}\biggl(\Att\bigl[F(Y_1)Y_{t-n}^2\bigr]-\Att[F(Y_1)]\cdot\Att\bigl[Y_1^2\bigr]\biggr)~,
\label{recFY}
\end{equation}
which establishes a recursive scheme to compute
$\Att\bigl[F(Y_1)Y_t^2\bigr]-\Att[F(Y_1)]\cdot\Att\bigl[Y_1^2\bigr]$.

Assuming $\alpha>4$, setting $F(y)=y^2$ in Eq.\ (\ref{recFY}), and
dividing by $\Att\bigl[|Y_1|^4\bigr]-\Att\bigl[|Y_1|^2\bigr]^2$, we
get a simple tool to evaluate $r_2^Y(t)$ for any $t$ and to study its
asymptotic behavior. In particular, as expected we find that 
if $2\le t\le M+1$ $r_2^Y(t)$ is
independent of the time and equal to
$1/(\alpha-1)$, while for $t>M+1$ 
\begin{equation}
r_2^Y(t)=\frac{1}{\alpha+M-2}\cdot\sum_{n=1}^{M}r_2^Y(t-n)~.
\label{r2Y}
\end{equation}

\begin{figure}
\centering
\psfrag{a}[ct][ct][2.]{$\alpha$}
\psfrag{lamb}[ct][ct][2.]{$\lambda$}
\psfrag{M=10}[ct][ct][1.5]{$M=10$}
\psfrag{M=20}[ct][ct][1.5]{$M=20$}
\psfrag{M=30}[ct][ct][1.5]{$M=30$}
\psfrag{M=40}[ct][ct][1.5]{$M=40$}
\psfrag{M=50}[ct][ct][1.5]{$M=50$}
\includegraphics[angle=-90,scale=0.5]{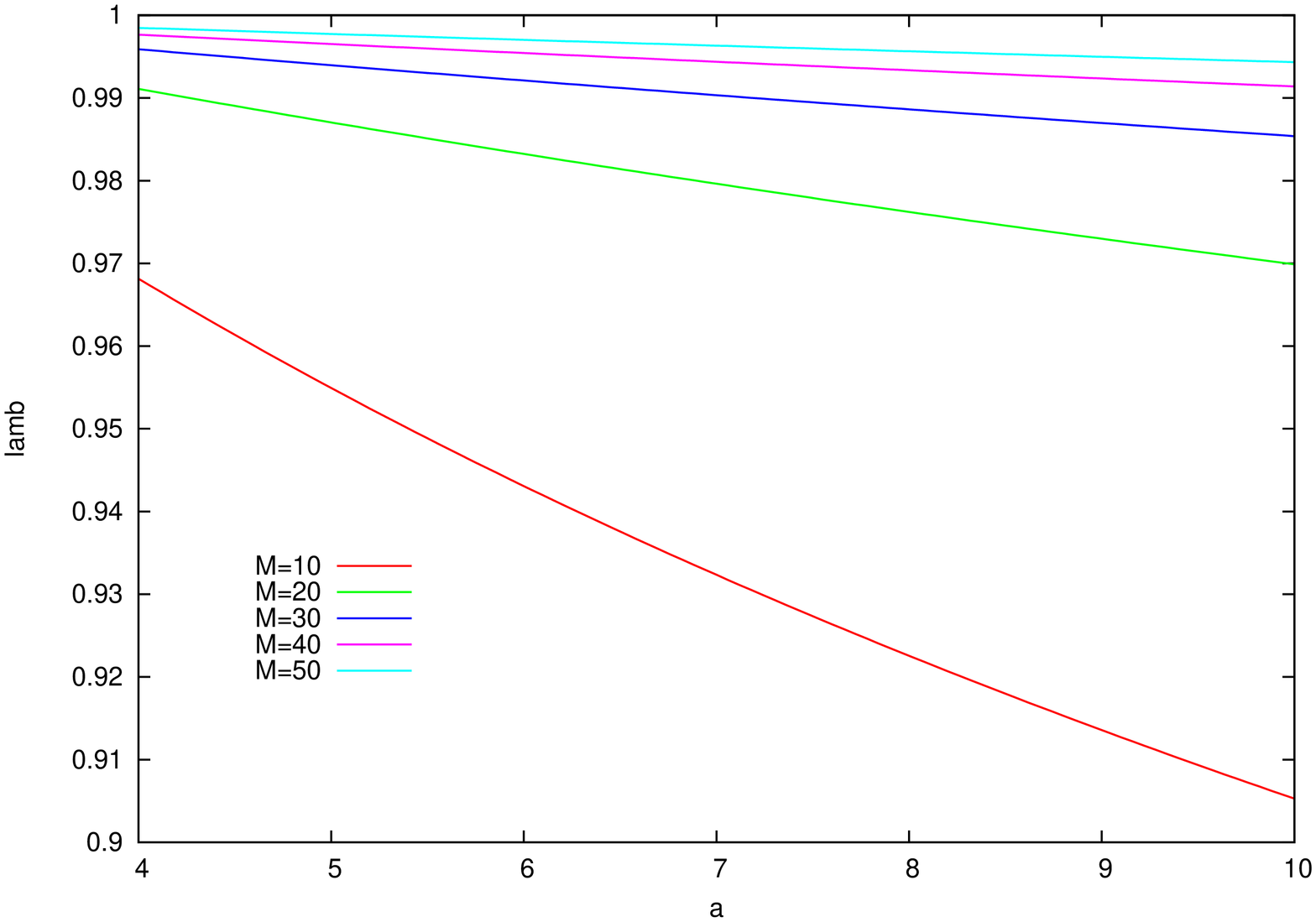}
\caption{Rate $\lambda$ as a function of $\alpha$ for different values
  of $M$. \label{fig:lamb}}
\end{figure}

In order to elucidate the asymptotic decay of $r_2^Y$, 
it is interesting to consider the
function
\begin{equation}
F(x)\equiv\frac{1}{\alpha+M-2}\cdot\sum_{n=1}^{M}\frac{1}{x^n}
\end{equation}
for positive $x$.  This is a strictly decreasing
positive continuous function which diverges to infinity when $x\to
0^+$ and goes to zero when $x\to +\infty$. Thus, there exists a unique
positive $\lambda$ such that $F(\lambda)=1$, which is smaller than 1
since $F(1)=M/(M+\alpha-2)<1$. The interest of this is that from
Eq.\ (\ref{r2Y}) we have that $r_2^Y$ decays exponentially fast with
rate $-\ln\lambda$.  Fig.\ \ref{fig:lamb} shows $\lambda$ vs.
$\alpha$ for different values of the memory $M$. Not surprisingly, the
correlations have a slower decay at higher values of $M$. Also, the
decay rate is minimum when $\alpha$ goes towards the lower limit value
of 4, namely when the distribution of $Y_1$ displays the most
pronounced tails.
To show that $\lambda$ really rules the decay of $r_2^Y$, we prove by
induction that
\begin{equation}
\frac{\lambda^{t-2}}{\alpha-1}\le r_2^Y(t)\le \frac{\lambda^{t-M-1}}{\alpha-1}
\label{br}
\end{equation}
if $t\ge 2$. These bounds then entail that
\begin{equation}
\lim_{t\to\infty}\frac{1}{t}\ln r_2^Y(t)=\ln\lambda~.
\end{equation}
We focus on the second inequality only, as the first can be treated
with the same arguments. Since $\lambda<1$ we have that
$\lambda^{t-M-1}/(\alpha-1)\ge 1/(\alpha-1)=r_2^Y(t)$ for $t$ in
between 2 and $M+1$. Fixing then $t>M+1$ and assuming that the
inequality holds up to $t-1$, we see that
\begin{equation}
r_2^Y(t)\le \frac{1}{\alpha+M-2}\cdot\sum_{n=1}^{M}\frac{\lambda^{t-n-M-1}}{\alpha-1}
=\frac{\lambda^{t-M-1}}{\alpha-1}\cdot F(\lambda)=\frac{\lambda^{t-M-1}}{\alpha-1}~.
\end{equation}

\subsection{The autocorrelation $r_q^X$}
We now bring together the above results to discuss the behavior of 
$r_q^X$. Eq.\ (\ref{rX}) tell us that
\begin{equation}
r_q^X(t)=u_q+v_q~r_q^{a_I}(t).
\label{rqXshort}
\end{equation}
If $2\le t\le M+1$ the coefficients $u_q$ and $v_q$ are independent of time:
\begin{eqnarray}
u_q&\equiv&\frac{\Att\bigl[|Y_1|^q|Y_2|^q\bigr]-\Att\bigl[|Y_1|^q\bigr]^2}
{\Att\bigl[a^{2q}_{I_1}\bigr]\Att\bigl[|Y_1|^{2q}\bigr]-\Att\bigl[a^q_{I_1}\bigr]^2\Att\bigl[|Y_1|^q\bigr]^2}\cdot \Att\bigl[a^q_{I_1}\bigr]^2,
\\
v_q&\equiv&\frac{\Att\bigl[a^{2q}_{I_1}\bigr]-\Att\bigl[a^q_{I_1}\bigr]^2}
{\Att\bigl[a^{2q}_{I_1}\bigr]\Att\bigl[|Y_1|^{2q}\bigr]-\Att\bigl[a^q_{I_1}\bigr]^2\Att\bigl[|Y_1|^q\bigr]^2}\cdot\Att\bigl[|Y_1|^q|Y_2|^q\bigr]~.
\end{eqnarray}
Thus, at short time scales the autocorrelation $r_q^X$ entirely
inherits the time dependence of $r_q^{a_I}$, as we mentioned in the
Main Text. Notice that $u_q=0$ when the endogenous $\Y$ process reduces to
a sequence of independent variables and in such a case
Eq.\ (\ref{rqXshort}) holds for any $t$.  More in general, for any $t$
we can rewrite Eq.\ (\ref{rX}) as
\begin{equation}
r_q^X(t)=u_q(t)~r_q^Y(t)+v_q~r_q^{a_I}(t)
\end{equation}
where we have again used the letters $u_q$ and $v_q$ but with a different
meaning. Here $u_q$ is the positive function of the time
\begin{equation}
u_q(t)\equiv\frac{\Att\bigl[|Y_1|^{2q}\bigr]-\Att\bigl[|Y_1|^q\bigr]^2}
{\Att\bigl[a^{2q}_{I_1}\bigr]\Att\bigl[|Y_1|^{2q}\bigr]-\Att\bigl[a^q_{I_1}\bigr]^2\Att\bigl[|Y_1|^q\bigr]^2}\cdot\Att\bigl[a^q_{I_1} a^q_{I_t}\bigr]
\end{equation}
and $v_q$ the positive coefficient
\begin{equation}
v_q\equiv\frac{\Att\bigl[a^{2q}_{I_1}\bigr]-\Att\bigl[a^q_{I_1}\bigr]^2}
{\Att\bigl[a^{2q}_{I_1}\bigr]\Att\bigl[|Y_1|^{2q}\bigr]-\Att\bigl[a^q_{I_1}\bigr]^2\Att\bigl[|Y_1|^q\bigr]^2}\cdot\Att\bigl[|Y_1|^q\bigr]^2~.
\end{equation}
Since
\begin{equation}
\lim_{t\to\infty}u_q(t)=\frac{\Att\bigl[|Y_1|^{2q}\bigr]-\Att\bigl[|Y_1|^q\bigr]^2}
{\Att\bigl[a^{2q}_{I_1}\bigr]\Att\bigl[|Y_1|^{2q}\bigr]-\Att\bigl[a^q_{I_1}\bigr]^2\Att\bigl[|Y_1|^q\bigr]^2}\cdot\Att\bigl[a^q_{I_1}\bigr]^2>0~,
\end{equation}
when $\rho$ is given by Eq.\ (\rhopaper) of the Main Text
(for which we know the behavior of $r_q^Y$ in the instance $q=2$) we
find that
\begin{equation}
\lim_{t\to\infty}\frac{1}{t}\ln r_2^X(t)=\ln\max\{\lambda,1-\nu\}~.
\label{raterX}
\end{equation}
Thus, within this setting the autocorrelation of the observed process decays
exponentially fast in the long time limit. 
The slowest between the relaxation
rates of $r_2^Y$ and
$r_2^{a_I}$ determine the one of $r_q^X$.

\section*{Acknowledgments}
This work is supported by ``Fondazione Cassa di Risparmio di Padova e
Rovigo'' within the 2008-2009 ``Progetti di Eccellenza'' program.


\begin{thebibliography}{99}
\bibitem{note_andreoli}
This version of the model results to be the discrete-time
  counterpart of the one considered in Ref.\ \cite{andreoli_1} after
  one keeps a constant average volatility.
\end{thebibliography}
\end{document}